\documentclass[pra,amsmath,amssymb,floatfix,twocolumn]{revtex4}

\usepackage{graphicx}
\usepackage{subfigure}
\usepackage{dcolumn}
\usepackage{verbatim}
\usepackage{epstopdf}

\def\dd{\textrm{d}}
\def\Tr{\textrm}
\def\Bf{\boldsymbol}
\def\rv{\Bf{r}}

\def\Av{\Bf{A}}

\def\nv{\Bf{\nabla}}
\def\qn{\Bf{n}}
\def\xh{\hat{\Bf{x}}}

\def\la{\langle}

\begin{document}
\title{The phase diagram of two-dimensional fast-rotating ultra-cold fermionic atoms near unitarity}

\author{Predrag Nikoli\'c}
\affiliation{Department of Physics, Rice University, Houston, TX 77005}

\begin{abstract}

By analyzing vortex lattices, re-entrant Cooper pairing and Fulde-Ferrell-Larkin-Ovchinnikov (FFLO) states in a single theoretical framework we explore how vortices and spin textures join to protect superconductivity against large magnetic fields. We use a rapidly rotating ultra-cold gas of fermionic atoms near unitarity as a model system amenable to experimental exploration, and discover a hierarchy of spin-polarized and FFLO phases in which a metal or a band-insulator of unpaired particles coexists with a spatially modulated superfluid hosting a vortex lattice. Quantum fluctuations can transform these phases into strongly correlated ``vortex liquid'' metals and insulators respectively. We argue that vortex lattices significantly enhance the stability of FFLO states and discuss prospects for observing these states in cold atom experiments.

\end{abstract}

\date{\today}

\maketitle

\section{Introduction}

Theoretically anticipated mechanisms for the survival of superconductivity in large magnetic fields include vortex lattice formation \cite{Abrikosov1957}, re-entrant superconductivity (pairing of Landau-localized electrons well above $H_{c2}$) \cite{rasolt92, Rieck1990, Norman1990}, and FFLO states (partial yielding to Zeeman effect involving spontaneous spatial modulations) \cite{Fulde1964, Larkin1964}. However, only the first mechanism has been unambiguously observed in nature \cite{Essmann1967}, and we are fortunate that a new frontier has been opened in atomic physics to observe and study these phenomena.

Remarkable tunability and qualitative similarities to electronic systems make trapped ultra-cold alkali atoms ideal for exploring a wide range of phenomena related to superconductivity \cite{Bloch08p885}. The orbital effect of magnetic field on electrons can be simulated by rotating atomic clouds, allowing Coriolis forces in the rotating frame to produce quantized vortices or quantum Hall effect. Paramagnetic or Zeeman effect can be simulated by trapping unequal numbers of atoms in the two lowest-energy hyperfine states, which have the same dynamics as electron spin and will be referred to as ``spin'' in this paper. The Zeeman and orbital effects can be independently controlled in cold atoms. The strength of attractive interactions between atoms, responsible for superfluidity, is also easily tuned using Feshbach resonances.

The competition between superfluidity and Zeeman effect is usually resolved by forming either an unpolarized superfluid of Cooper pairs, or a spin-polarized normal state \cite{Zwierlein2006, Shin06, Partridge2006, Partridge06}. The elusive FFLO state is a compromise in which spatially alternating superfluidity and spin-polarization coexist. FFLO states have been studied both in high energy and condensed matter physics \cite{Casalbuoni2004}. Three-dimensional \cite{Gruenberg1966, Buzdin1996, Shim2006} and two-dimensional \cite{Yang2004, Shimahara1998, Houzet2002, Shimahara1998b, Klein2004, Matsuda2007} FFLO states have been analyzed in superconductors, heavy-fermion metals and cold atom gases. This work is partially motivated by the search for FFLO states in layered materials with tilted magnetic field.

In this paper we examine the relationship between FFLO phenomena, re-entrant superfluidity, integer quantum Hall states and strongly correlated insulators of general interest in unconventional superconductivity, using a previously unexplored rotating cold-atom perspective (for a review of past research, mostly on rotating bosons or three-dimensional fermionic systems, see Ref.\cite{Cooper2008} and references therein). We determine superfluid order parameter by minimizing free energy of rotated fermions in two dimensions near unitarity. This reveals the first order transitions expected from Zeeman effect and provides detailed insight into quasiparticle spectra. We present a rich zero-temperature phase diagram of competing states in the background of quantum Hall effect, which is beyond reach of earlier studies based on pairing instability. We also discuss manifestations of this phase diagram in hypothetical time-of-flight experiments, which take form analogous to ``quantum oscillations'' in condensed matter systems (periodicity of various quantities due to Landau quantization of the quasiparticle density of states, such as Shubnikov - de Hass effect).

The quantitative applicability of the following universal theory to ultra-cold fermionic atoms near the Feshbach resonance stems from the unitarity limit \cite{nikolic:033608}. The scattering length of two-body interactions diverges at the resonance, so that the microscopic details of interactions become irrelevant and all properties of the many-body system become functions of only dimensionless ratios of externally controlled parameters. Such scaling laws are protected inside all superfluid phases, but may break down due to rotation in two dimensions when strongly correlated ``vortex liquid'' states are formed \cite{nikolic:144507}.

Most results of this paper are derived from a Bogoliubov-de Gennes (BdG) model which arises as a saddle-point (or mean-field) approximation in the quantum field theory of neutral fermionic particles near unitarity (written in section \ref{QFluct} and discussed in Ref.\cite{nikolic:144507}). This approximation becomes quantitatively accurate far from the Feshbach resonance in the BCS (Bardeen, Cooper, Schrieffer) limit, and more so at larger densities, but remains extremely useful even in the numerically accessible regimes for elucidating various qualitative features which may be observed in experiments. Rotation at angular velocity $\omega$ is described by a static gauge field $\Av$ in the rotating frame which satisfies $\nv\times\Av=2m\omega\hat{\Bf{z}}$, where $m$ is atomic mass. We assume that centrifugal forces are almost exactly cancelled by the potential which traps atoms. This automatically puts us in the deep quantum limit because the trapped gas expands to a state with very low density. We will work in grand-canonical ensemble and control particle densities by chemical potential $\mu$ and Zeeman field $h$, which is the best choice for predicting observable ``quantum oscillations'' despite the direct experimental control of particle numbers. After presenting the qualitative pairing phase diagram and analyzing experimental prospects, we provide technical details of calculations and discuss the effects of quantum fluctuations.

\section{Pairing phase diagram: spin-polarized vortex lattices}\label{secMain}

In the absence of interactions all atoms would be localized in their Landau orbitals created by Coriolis forces in the rotating frame. Such localization facilitates Cooper pairing and allows re-entrant superfluidity to occur even in the deep quantum limit at arbitrarily high rotation rates \cite{rasolt92, Rieck1990, Norman1990}. Saddle-point approximation regards the superfluid as a macroscopic condensate of Cooper pairs into a single state given by the order parameter function:
\begin{equation}\label{OP}
\Phi(\rv) = \sum_n \sum_{l=0}^{n_v-1}
  \phi_{n,l} \sum_j 2^{\frac{1}{4}} \varphi_n \left(
  \rv \sqrt{\frac{4m\omega}{\hbar}} ~ ; ~ \frac{(l + n_v j)\delta q}{\sqrt{4m\hbar\omega}} \right)
\end{equation}
The amplitudes $\phi_{n,l}$ describe the occupation of bosonic Landau levels $n\in\lbrace 0,1,2\dots\rbrace$ at wavevectors $q_x = \textrm{integer} \times \delta q$ in $x$-direction. The Landau level wavefunctions of dimensionless coordinates $\Bf{\xi}=(\xi_x,\xi_y)$ are:
\begin{equation}\label{WF}
\varphi_n(\Bf{\xi} ; \eta) = \frac{1}{\sqrt{2^n n! \sqrt{\pi}}} e^{i\eta\xi_x}
  e^{-\frac{1}{2}(\xi_y+\eta)^2} H_n(\xi_y+\eta)
\end{equation}
in Landau gauge, where $H_n$ are Hermite polynomials. This realizes a vortex lattice with periods $\Delta x = 2\pi\hbar / \delta q$ and $\Delta y = n_v \delta q (4m\omega)^{-1}$ containing $n_v$ vortices per unit-cell. Hexagonal Abrikosov lattices have $n_v=2$, $\delta q = (4\pi\sqrt{3}m\hbar\omega)^{1/2}$, and $\phi_{n,1} = (-1)^n i\phi_{n,0}$.

The BdG Hamiltonian acting on Nambu spinors $\vert \psi_{\uparrow}^{\phantom{\dagger}} , \psi_{\downarrow}^{\dagger} \rangle$:
\begin{equation}\label{BdG}
H = \left(
  \begin{array}{cc}
     \frac{\lbrack-i\hbar\nv-\Av(\rv)\rbrack^2}{2m}-\mu-h & \Phi(\rv) \\
     \Phi^*(\rv) & -\frac{\lbrack-i\hbar\nv+\Av(\rv)\rbrack^2}{2m}+\mu-h
  \end{array}
\right) \nonumber
\end{equation}
determines the quantum-mechanical spectrum $E_{\qn}(\phi_{n,l})$ of fermionic quasiparticle excitations with quantum numbers $\qn$ in the condensate. We exactly numerically diagonalize this Hamiltonian in the basis of Landau orbitals. The energy spectrum $E_{\qn}=-h\pm\epsilon_{\qn}$ determines the free energy $\mathcal{F}(\phi_{n,l})$ in the saddle-point approximation \cite{L2000}:
\begin{eqnarray}\label{FreeEnergy}
&& \mathcal{F}(\phi_{n,l}) = - \sum_{\qn} \Bigl\lbrack
     k_B T \ln\left( 1+e^{-\beta(\epsilon_{\qn}(\phi_{n,l})+h)} \right) \\
&& ~~ + k_B T \ln\left( 1+e^{-\beta(\epsilon_{\qn}(\phi_{n,l})-h)} \right)
      + \epsilon_{\qn}(\phi_{n,l}) - \epsilon_{\qn}(0) \Bigr\rbrack
   \nonumber \\
&& ~~ + \left\lbrack \frac{m\nu}{4\pi} - \Pi_{0,0}(0,0)\Bigr\vert_{\mu=h=T=0} \right\rbrack \int\dd r |\Phi(\rv)|^2
  \ . \nonumber
\end{eqnarray}
Here, $\beta=(k_B T)^{-1}$, $k_B$ is Boltzmann constant, and $\nu=-a_z/a$ measures the strength of attractive interactions between particles through the two-body scattering length $a$ and a confinement scale $a_z$ in $z$-direction which renders the system two-dimensional. Therefore, $\nu$ is proportional to the detuning from the Feshbach resonance in experiments. The ultra-violet behavior is regulated by rotation, and the subtracted inverse pairing susceptibility $\Pi_{n_1,n_2}(i\omega,p_x)$ assures that $\nu=0$ corresponds to unitarity \cite{nikolic:144507}. The global minimum of $\mathcal{F}(\phi_{n,l})$ determines $\phi_{n,l}$ in the thermodynamic equilibrium.

\begin{figure}
\centering
\includegraphics[width=2.3in]{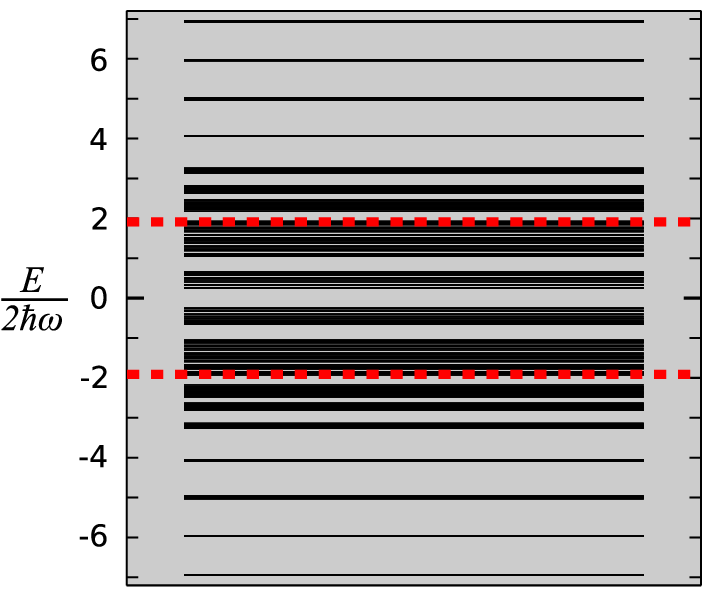}
\caption{\label{BdGspectrum}(color online) A typical spectrum of quasiparticle excitations in the superfluid state, taken at unitarity, $h=0$, $\mu=2.705\times2\hbar\omega$. Red dashed lines denote the pairing gap scale defined by ~(\ref{pgap}).}
\end{figure}

The BdG spectrum $E_{\qn}$ in the superfluid hosting a vortex lattice at $h=0$ has a particle-hole symmetric band-structure reminiscent of broadened Landau levels at high energies (see Fig.\ref{BdGspectrum}). The lowest quasiparticle bands are most affected by pairing and their local density of states is generally maximized inside vortex cores due to gapping in the surrounding bulk. Deep in the superfluid phase the character of these bands is derived from bound vortex core states \cite{Caroli1964, Sensarma2006} by inter-core quantum tunneling. The Zeeman field $h$ acts as a chemical potential for the BdG quasiparticles which breaks the particle-hole symmetry. At zero temperature all states with $E_{\qn}<h$ are occupied, and with $E_{\qn}>h$ are empty. There is a ``metal-insulator'' quantum phase transition each time $h$ crosses a band-edge \cite{SubirQPT}. The injected quasiparticles at energies $0<E_{\qn}<h$ carry spin polarization, spatially modulated by the vortex lattice and maximized near vortex cores (spatial spin distribution near a single vortex is discussed in Ref.\cite{Takahashi2006}). The ``metal-insulator'' transitions and polarization profiles provide routes, respectively, to energy and position resolved spectroscopy of quasiparticle states in vortex lattices.

\begin{figure*}
\subfiguretopcaptrue\centering
\def\subfigcapmargin{0.4in}
\subfigure[{}]{\includegraphics[height=2.6in]{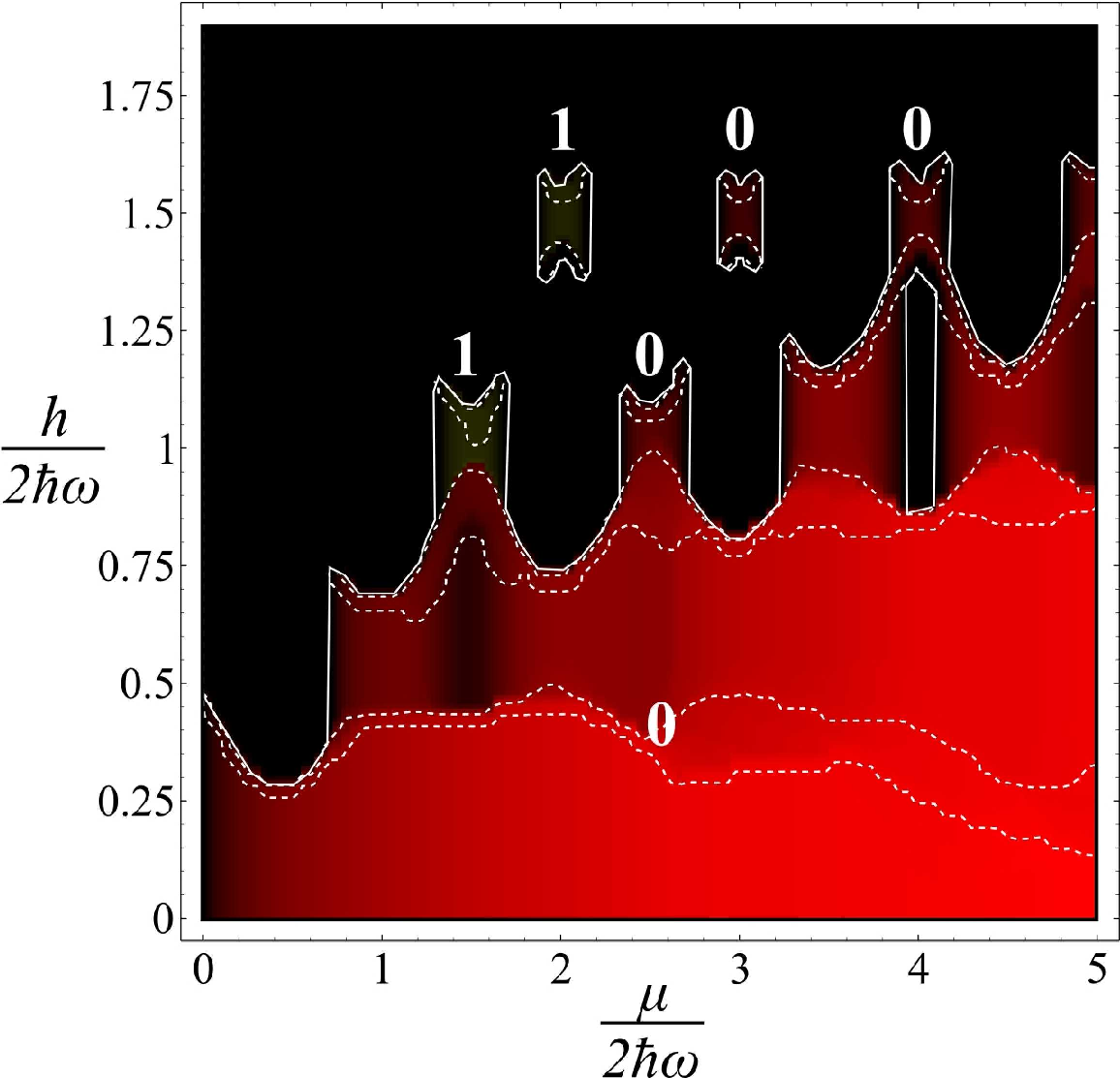}} \hskip 0.2in
\subfigure[{}]{\includegraphics[height=2.6in]{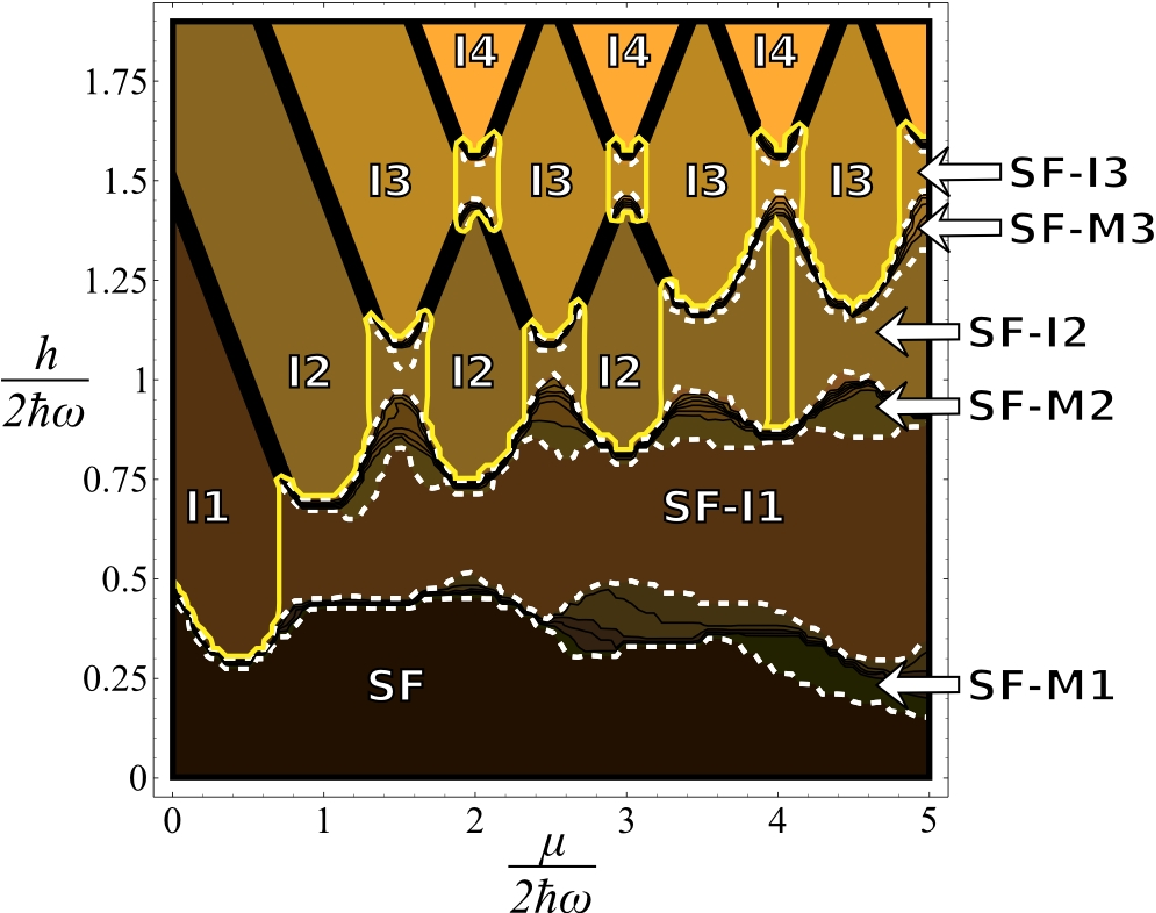}}
\subfigure[{}]{\includegraphics[height=2.6in]{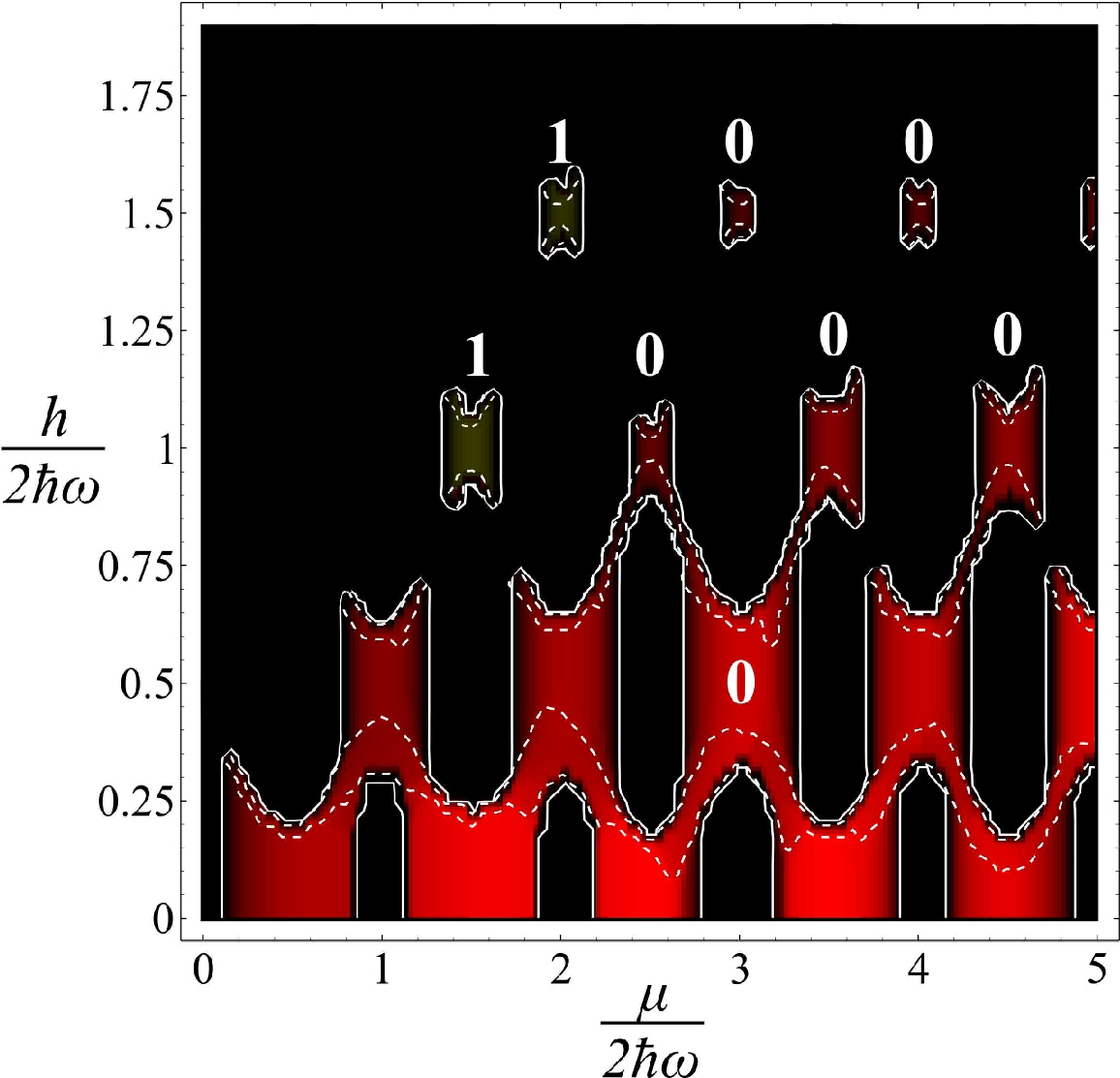}} \hskip 0.2in
\subfigure[{}]{\includegraphics[height=2.6in]{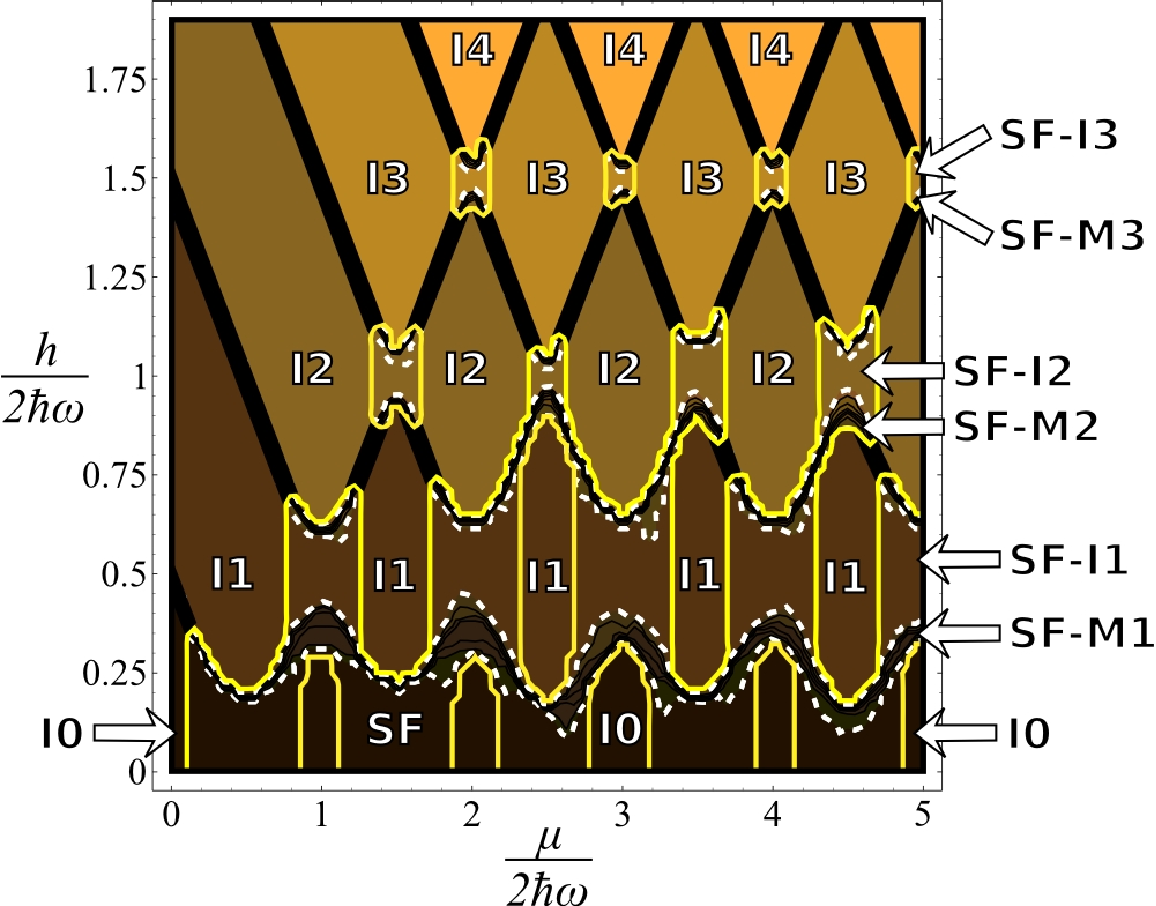}}
\caption{\label{PhaseDiagram}(color online) Quantum phase diagrams of fast-rotating fermionic atoms near unitarity, showing spatially averaged density of Cooper pairs $\rho_s$ and polarization $p=n_\uparrow-n_\downarrow$: (a) $\rho_s$ and (b) $p$ at unitarity; (c) $\rho_s$ and (d) $p$ in the BCS limit at $\nu=0.15$. The density plots of $\rho_s$ are color-coded by the dominant Landau level (LL) $n$ in the condensate, also indicated with numbers: red $n=0$, green $n=1$. The contour plots of $p$ emphasize superfluid-insulator transitions with bright full lines (straight vertical segments are $2^\textrm{nd}$ order, curved segments are $1^\textrm{st}$ order), metal-insulator transitions within the superfluid phase with bright dashed lines, and transitions between quantum Hall insulators (I) with thick-black lines. The superfluid phases (SF) host an Abrikosov vortex lattice, and may contain a coexisting metal (SF-M) or band-insulator (SF-I) of polarized unpaired fermions. The numbers in the phase labels indicate the highest populated quasiparticle band. Polarization is saturated at an integer number of ``spins'' per flux quantum in all SF, SF-I and I phases, while it is not quantized in SF-M phases.}
\end{figure*}

The formation of Cooper pairs in an integer quantum Hall insulator is rather similar to pairing instability in band-insulators created by optical lattice potentials. Chemical potential or detuning driven pairing transition is second order \cite{Chin2006, moon:230403, Burkov2009}, so the pairing gap can be much smaller than the quasiparticle band-gap, in this case the cyclotron gap $2\hbar\omega$. Here we shall define a single energy scale which represents the order parameter:
\begin{equation}\label{pgap}
|\varphi| = \left\lbrack \sum_n \sum_{l=0}^{n_v-1} |\phi_{n,l}|^2 \right\rbrack^{\frac{1}{2}} =
  \left\lbrack \frac{n_v \delta q}{A\sqrt{8m\hbar\omega}} \int \dd^2 r |\Phi(\rv)|^2 \right\rbrack^{\frac{1}{2}}
\end{equation}
where $A$ is the system area. This ``pairing gap'' is related to spatially averaged superfluid density $\rho_s\propto |\varphi|^2$ through an unknown microscopic cut-off scale.

\begin{figure*}
\subfigure[{}]{\includegraphics[width=2.3in]{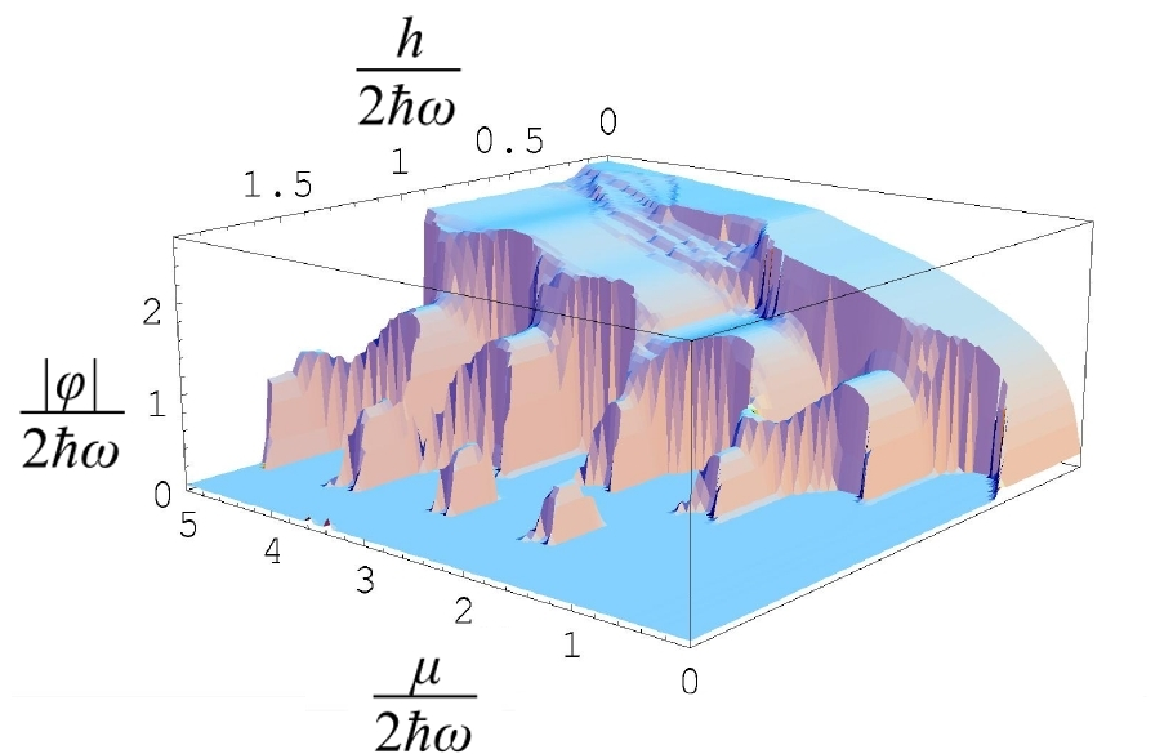}}
\subfigure[{}]{\includegraphics[width=2.3in]{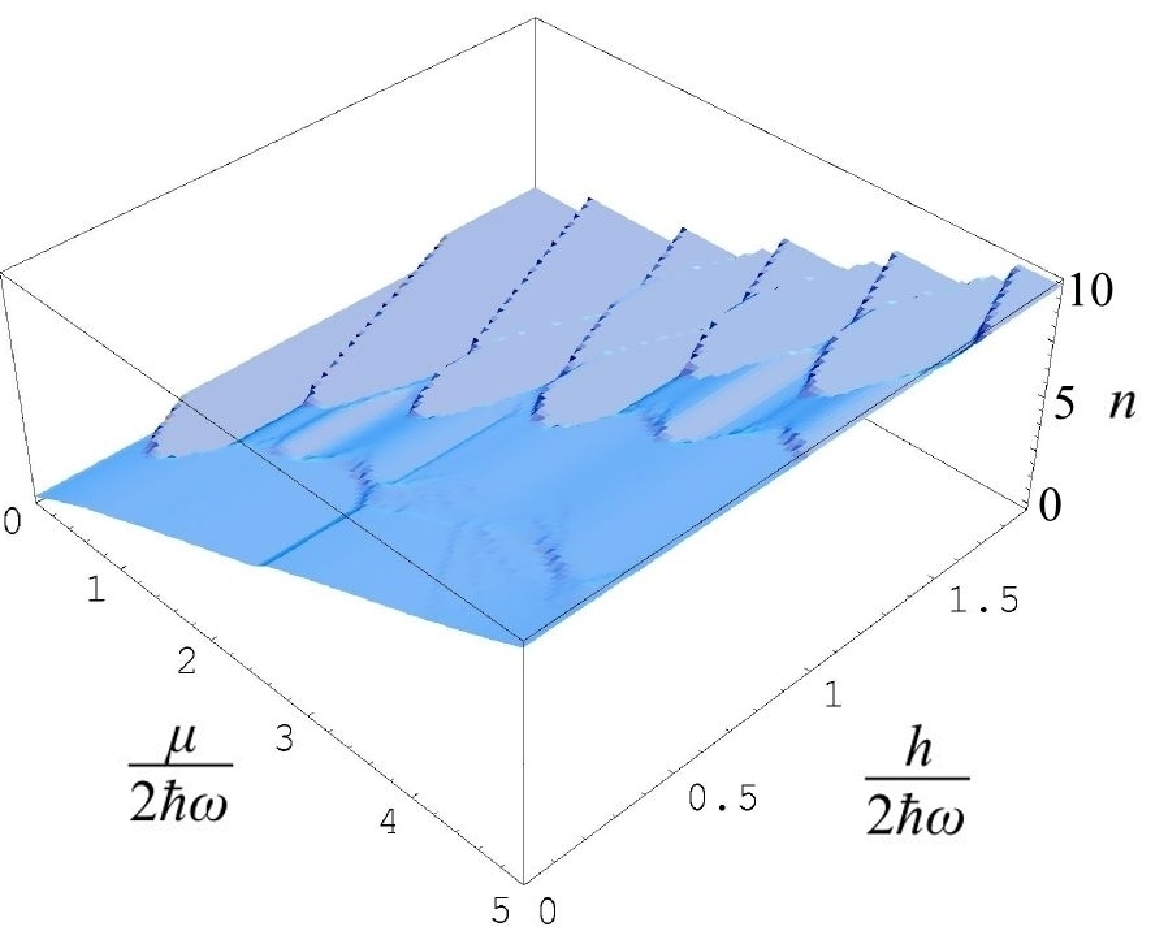}}
\subfigure[{}]{\includegraphics[width=2.3in]{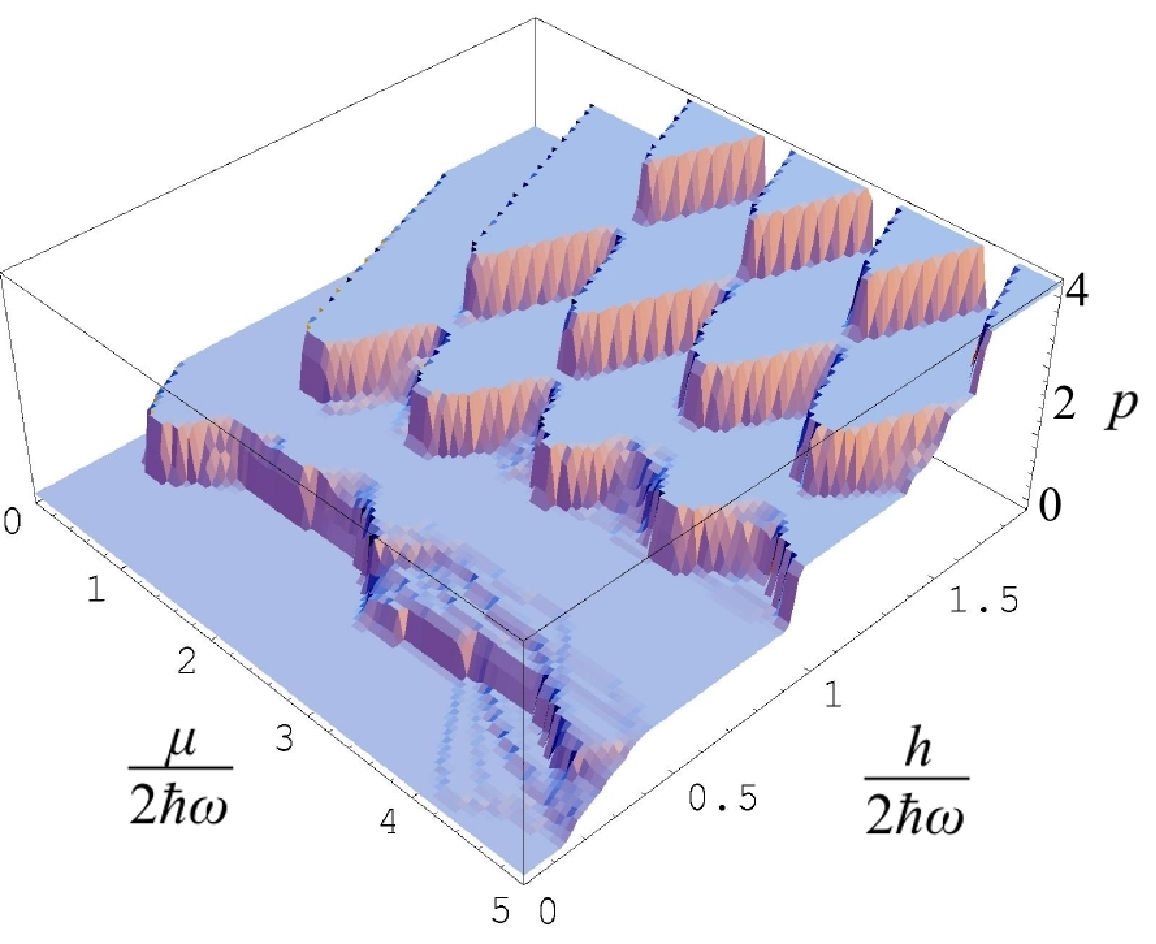}}
\caption{\label{phdiag3d}(color online) Three-dimensional plots of spatially averaged (a) pairing gap, (b) total particle density and (c) spin-polarization at unitarity, $T=0$. Density and polarization are expressed as a number of atoms or spins per flux quantum. This is the same data as shown in Fig.\ref{PhaseDiagram}(a,b).}
\end{figure*}

\begin{figure}
\centering
\includegraphics[width=2.5in]{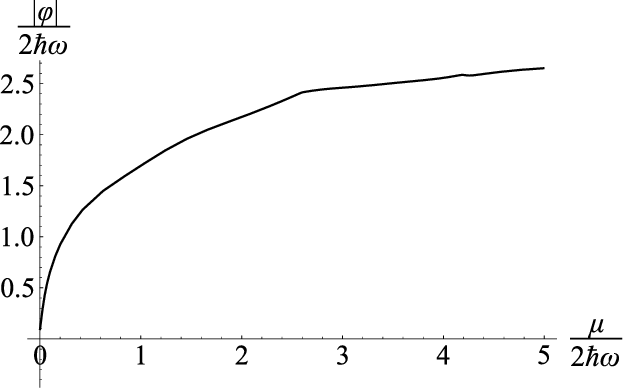}
\caption{\label{PairingGap}Pairing gap $|\varphi|$ from Eq.(\ref{pgap}) at $h=0$, $T=0$ and unitarity in the units of cyclotron gap $2\hbar\omega$.}
\end{figure}

We numerically minimized free energy ~(\ref{FreeEnergy}) using a fermionic basis of 72 quasi-periodic states in each of the 50 lowest Landau levels. The global minimum was calculated among order parameters within the lowest five bosonic Landau levels and $n_v=2$, $\delta q = (4\pi\sqrt{3}m\hbar\omega)^{1/2}$, allowing a variety of lattice structures besides the hexagonal. We use grand-canonical ensemble in order to obtain both the simplest presentation of main physical points and the simplest explanation of experimental signatures. The main features of the $T=0$ phase diagram are displayed in Fig.\ref{PhaseDiagram}, and additional plots at unitarity are shown in Fig.\ref{phdiag3d} and Fig.\ref{PairingGap} on the same data to better convey a variety of features.

The superfluid is divided into a hierarchy of phases with either gradually varying polarization (SF-M), or polarization saturated at an integer number of spins per vortex (SF-I). The unpaired polarized atoms form a Fermi surface in the former case, and a band-insulator in the latter case. The superfluid order parameter changes abruptly across some ``metal-insulator'' transitions, making them first order, while in other cases the present numerical calculation could not find such abrupt changes (they might exist in a larger space of order parameters than the one explored). This is best seen in Fig.\ref{phdiag3d}. First order transitions are very important since their manifestations are the most visible sharp features in experiments which can persist up to finite critical temperatures. Note that without jumps in the order parameter the $T=0$ ``metal-insulator'' transitions would broaden into crossovers at finite temperatures. Previous studies did not put emphasis on these features because they did not solve for the order parameter in the deep quantum limit. The $T=0$ phase diagram contains a sequence of quantum tri-critical points where the transitions between atomic quantum Hall insulators meet the first and second order superfluid-insulator transitions. The superfluid boundary also retains its pair-breaking first order transitions below some finite temperatures.

\begin{figure}
\includegraphics[height=2.6in]{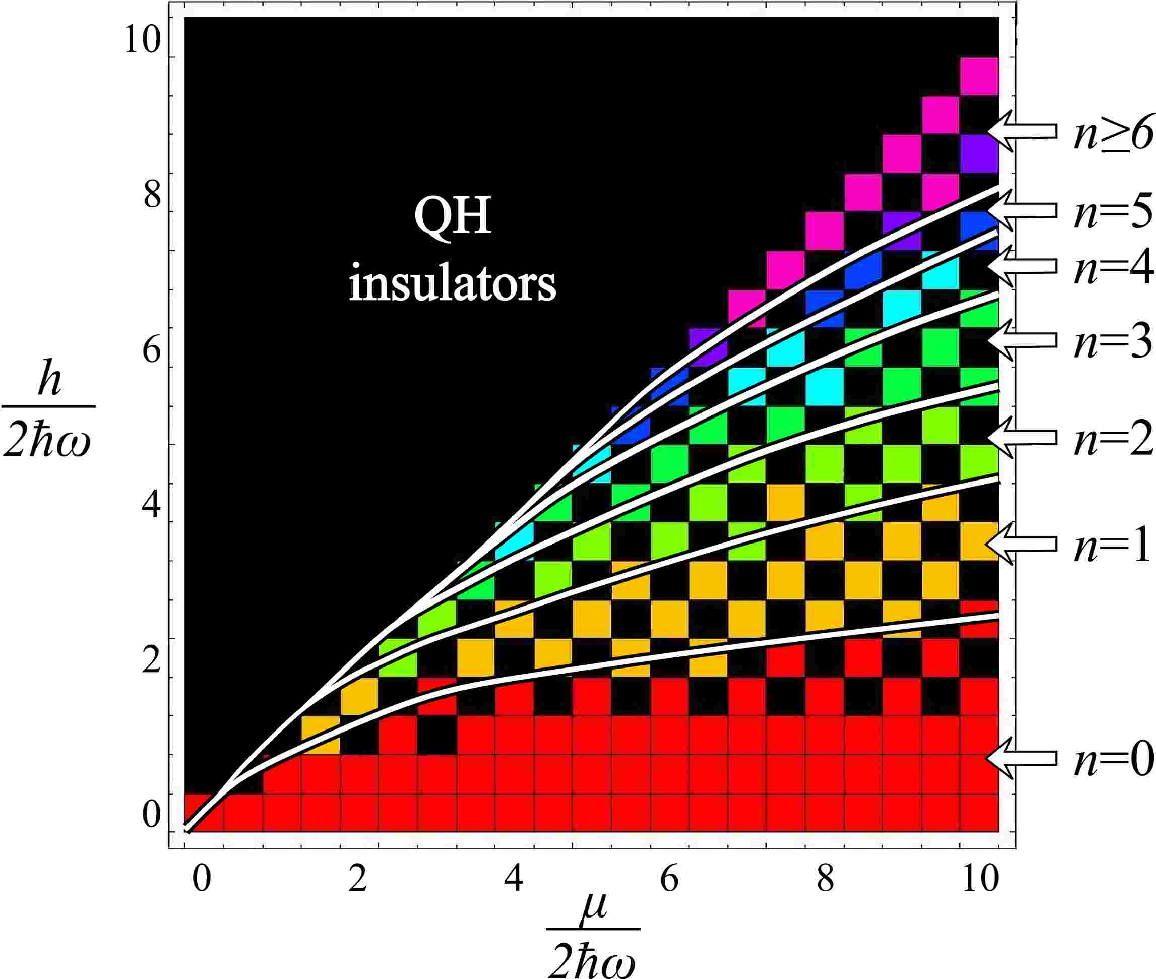}
\caption{\label{PhaseDiagramFFLO}(color online) Low-resolution color-coded map of FFLO states formed by condensation into higher bosonic Landau levels $n$. White lines are guides for the eye.}
\end{figure}

\begin{figure}
\subfiguretopcaptrue\centering
\subfigure[{}]{\includegraphics[height=0.6in]{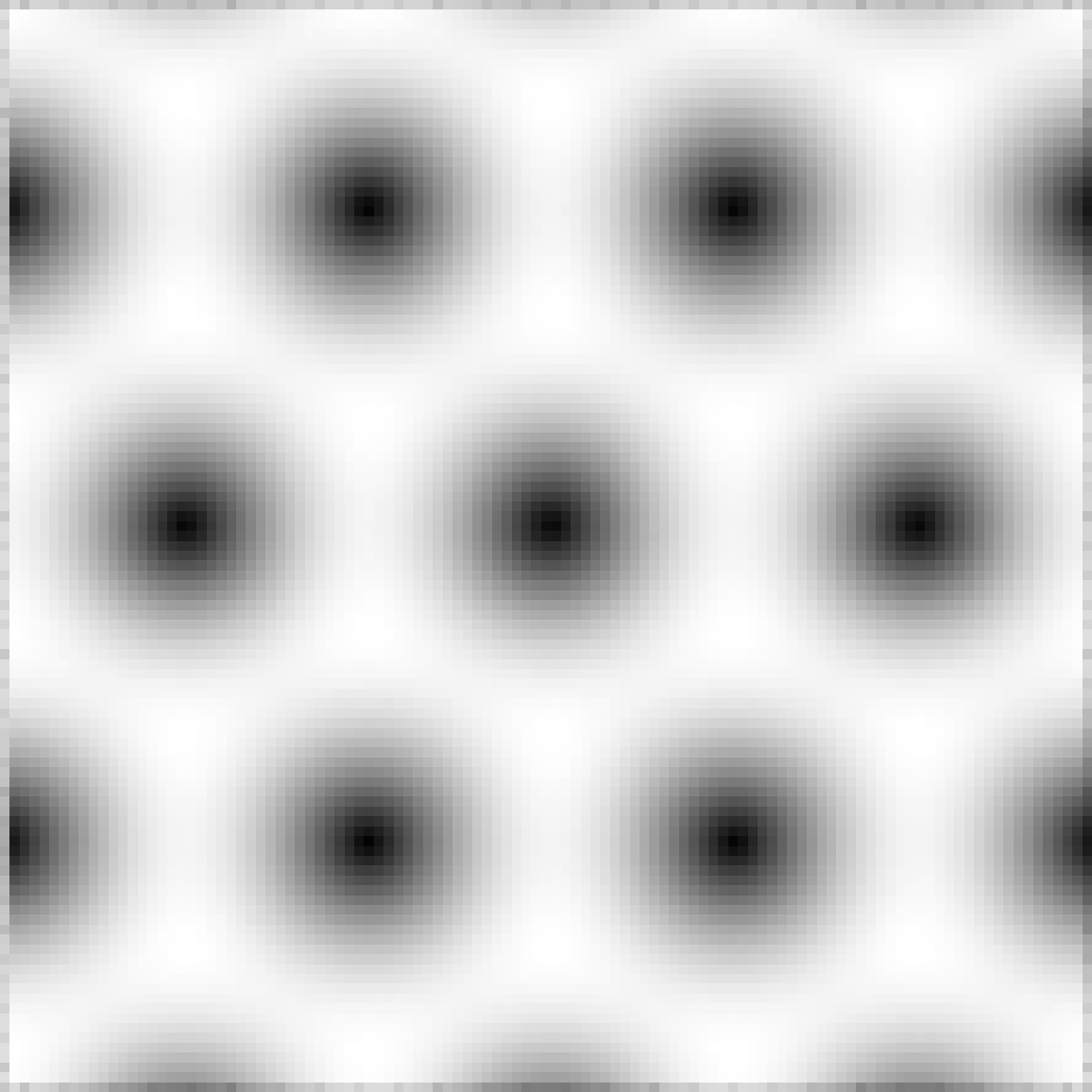}}
\subfigure[{}]{\includegraphics[height=0.6in]{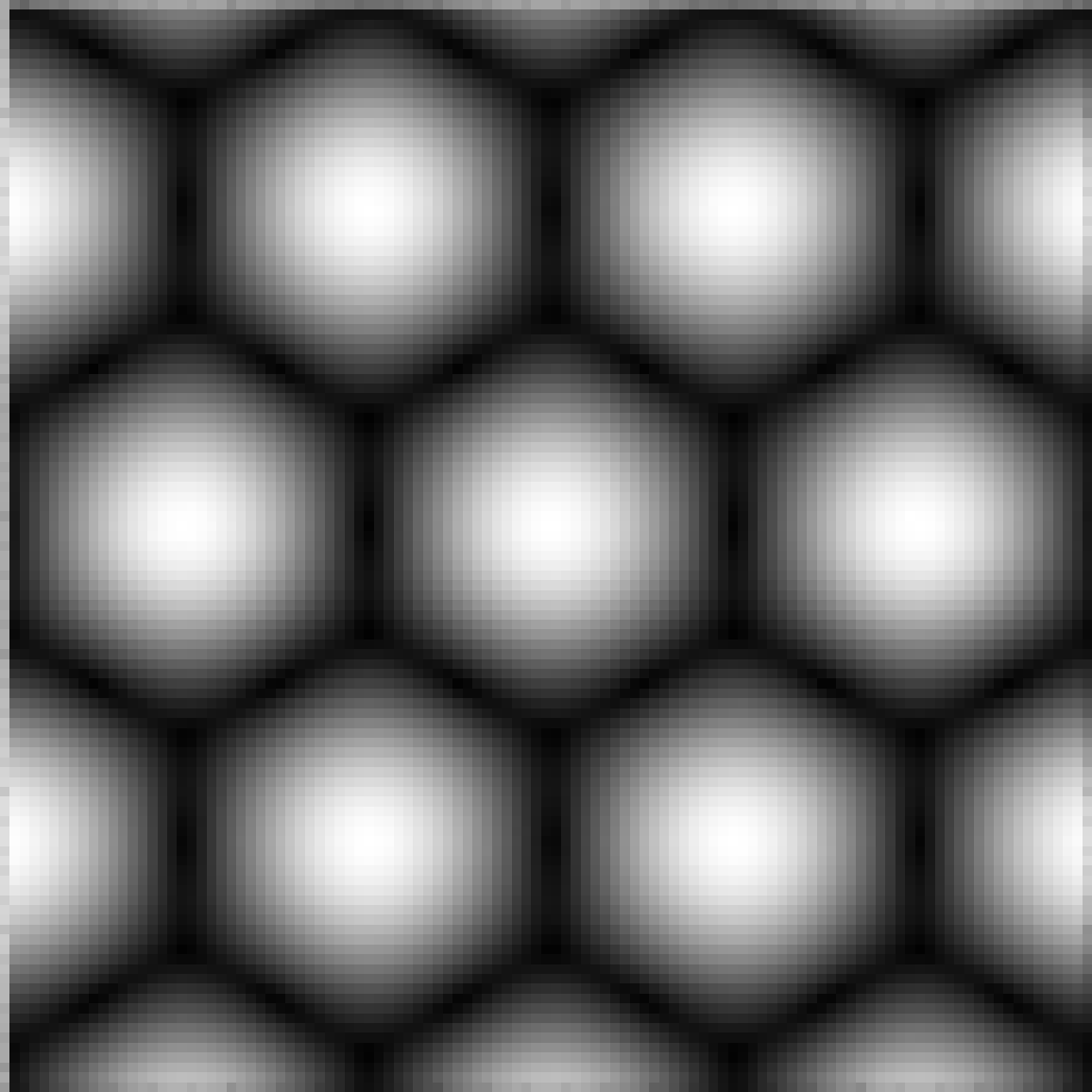}}
\subfigure[{}]{\includegraphics[height=0.6in]{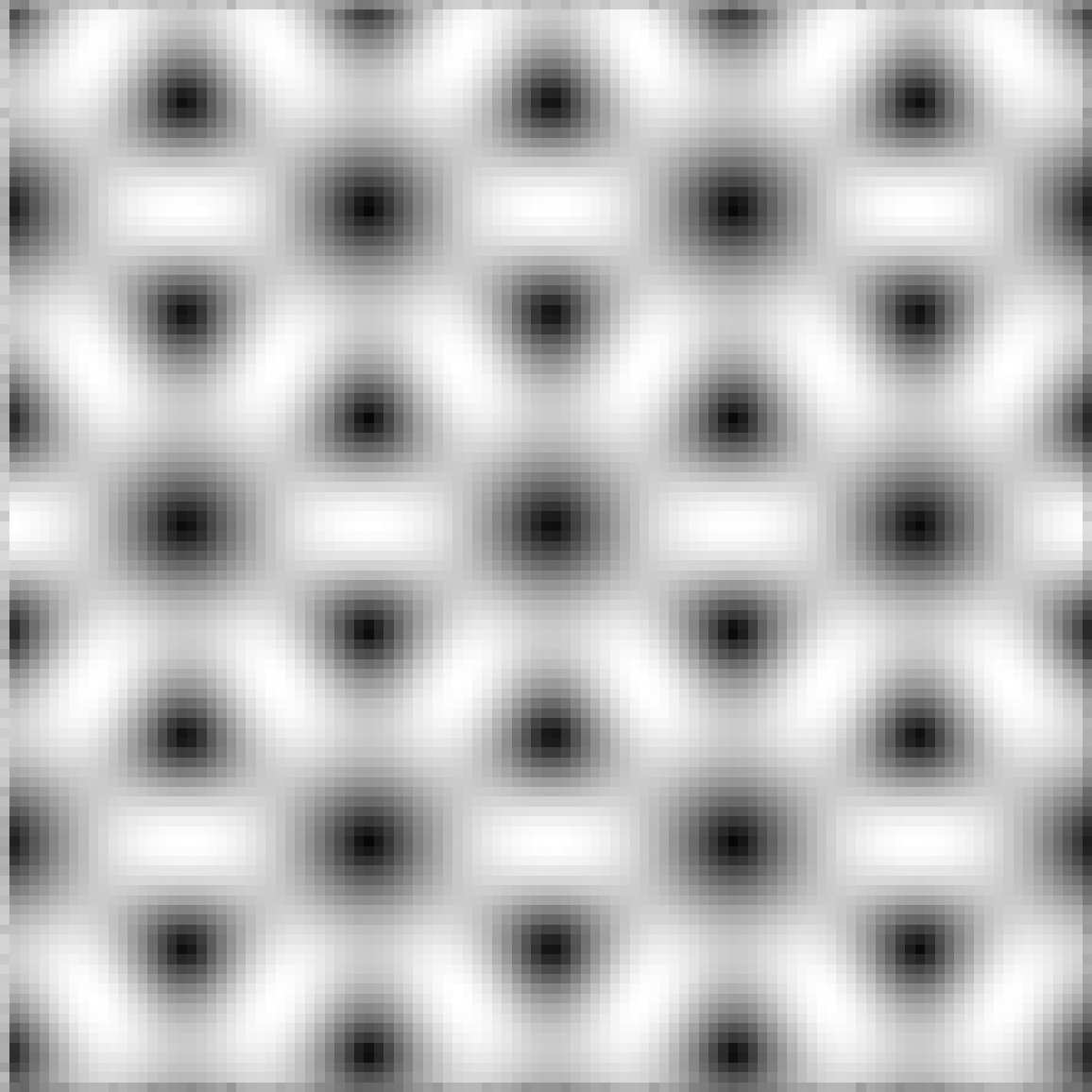}}
\subfigure[{}]{\includegraphics[height=0.6in]{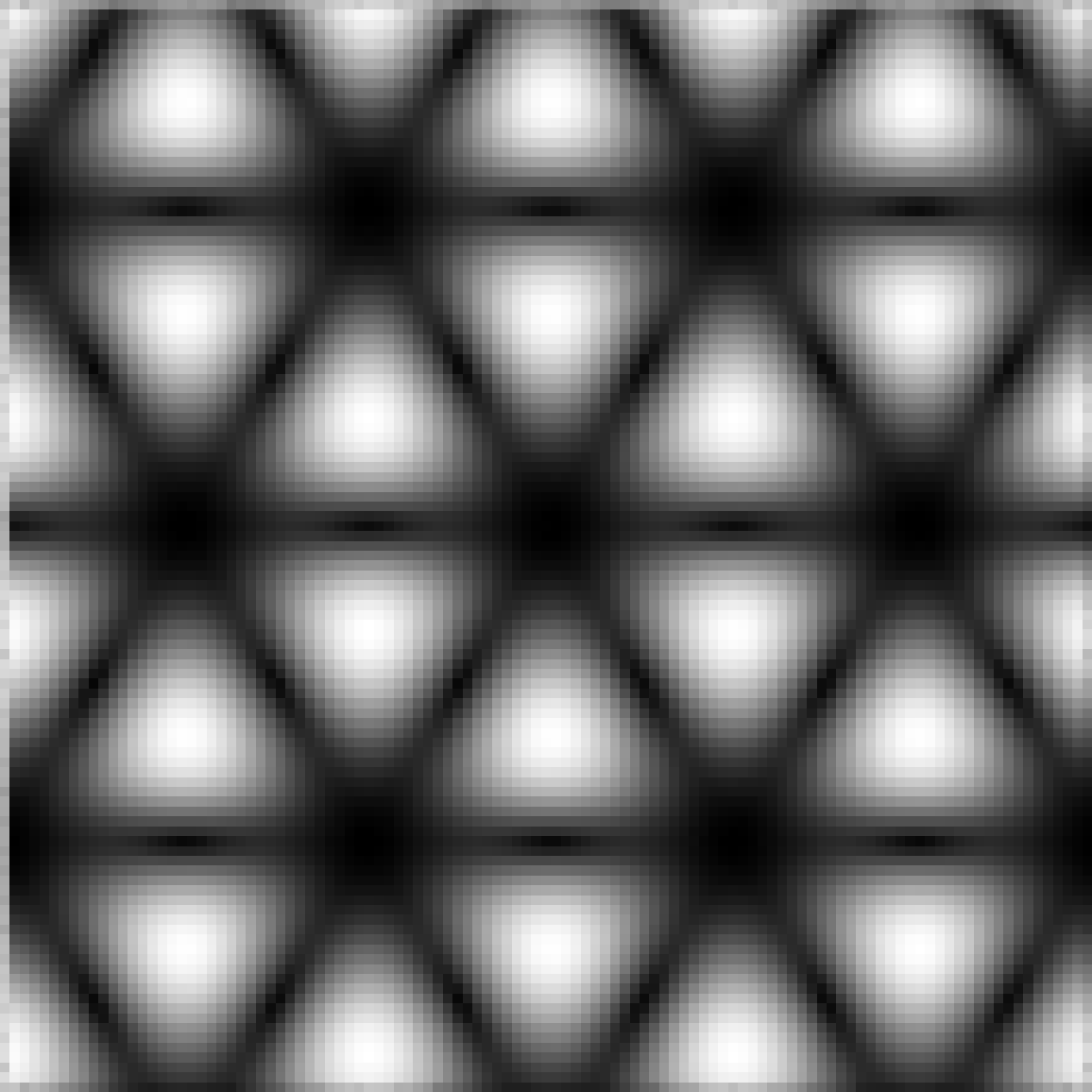}}
\subfigure[{}]{\includegraphics[height=0.6in]{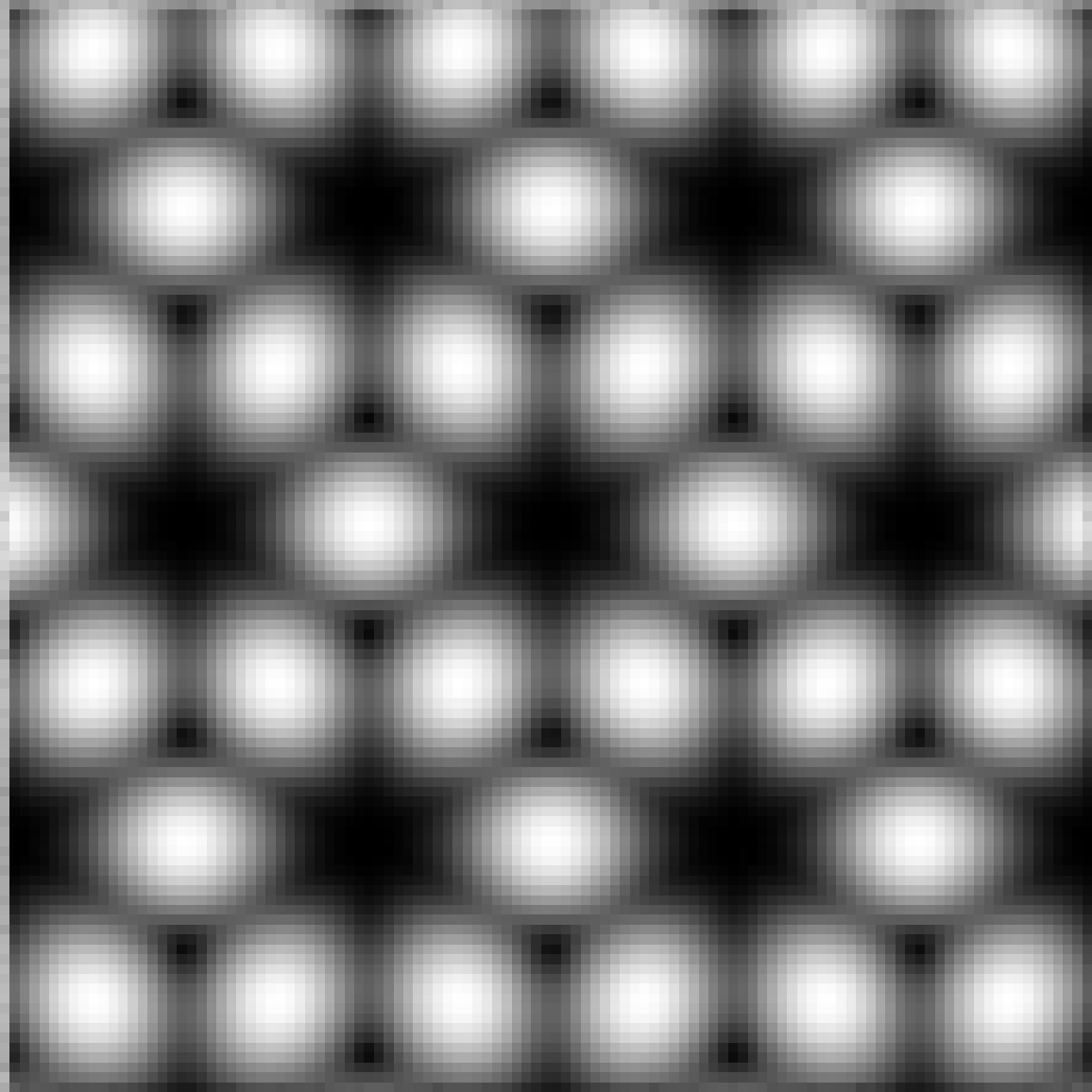}}
\caption{\label{VortexLattice}Density plots of hexagonal vortex lattices for condensates in (a) $n=0$, (b) $n=1$, (c) $n=2$, (d) $n=3$, and (e) $n=4$ Landau level. Brightness is normalized separately in each plot.}
\end{figure}

Zeeman effect can produce superfluids condensed in higher ($n \ge 1$) bosonic Landau levels \cite{Matsuda2007}. These are FFLO states in the presence of a vortex lattice. Typically, one Landau level is much more populated than the other ones, and the map of FFLO states is shown in Fig.\ref{PhaseDiagramFFLO}. Higher Landau levels are favored at large $h$ because they introduce $n$ additional vortex-antivortex pairs per vortex and hence have larger regions of suppressed superfluidity where local spin polarization can congregate. The presence of extra vortex-antivortex pairs spatially modifies superfluid density and allows a simple visual determination of the dominant bosonic Landau level as demonstrated in Fig.\ref{VortexLattice}.

Vortex lattice structural transitions within the same Landau level are also found in the saddle point approximation due to complicated inter-vortex forces mediated by fermionic degrees of freedom \cite{Stojanovic2007}. We do not show these structural transitions because quantum fluctuations of the order parameter generate familiar logarithmic ``Coulomb'' forces between vortices (known from Landau-Ginzburg theory) which dominate at low densities and act to stabilize the hexagonal Abrikosov lattice if it occurs in the lowest Landau level.

\begin{figure*}
\subfiguretopcaptrue\centering
\subfigure[{}]{\includegraphics[height=2.6in]{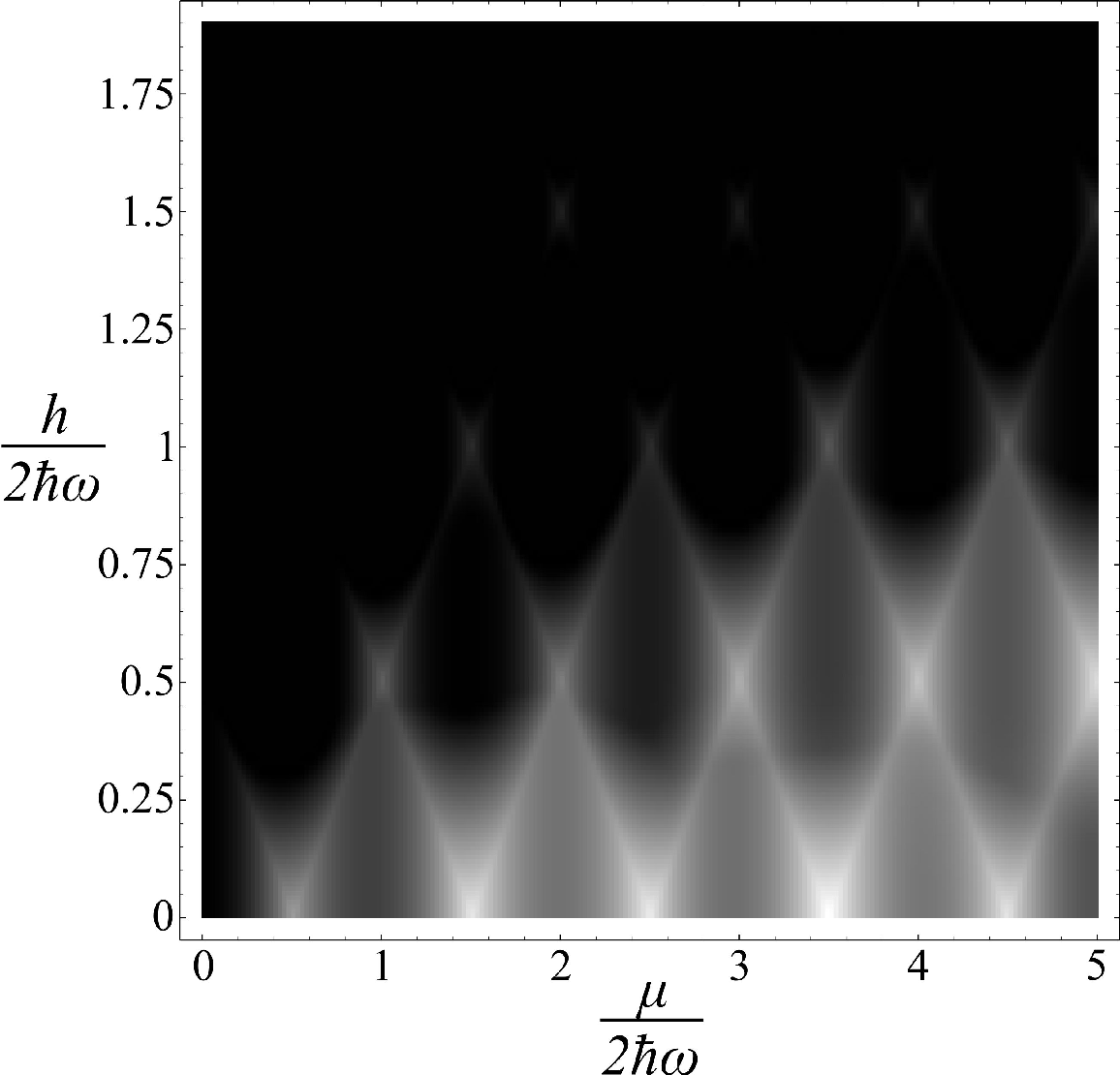}} \hskip 0.2in
\subfigure[{}]{\includegraphics[height=2.6in]{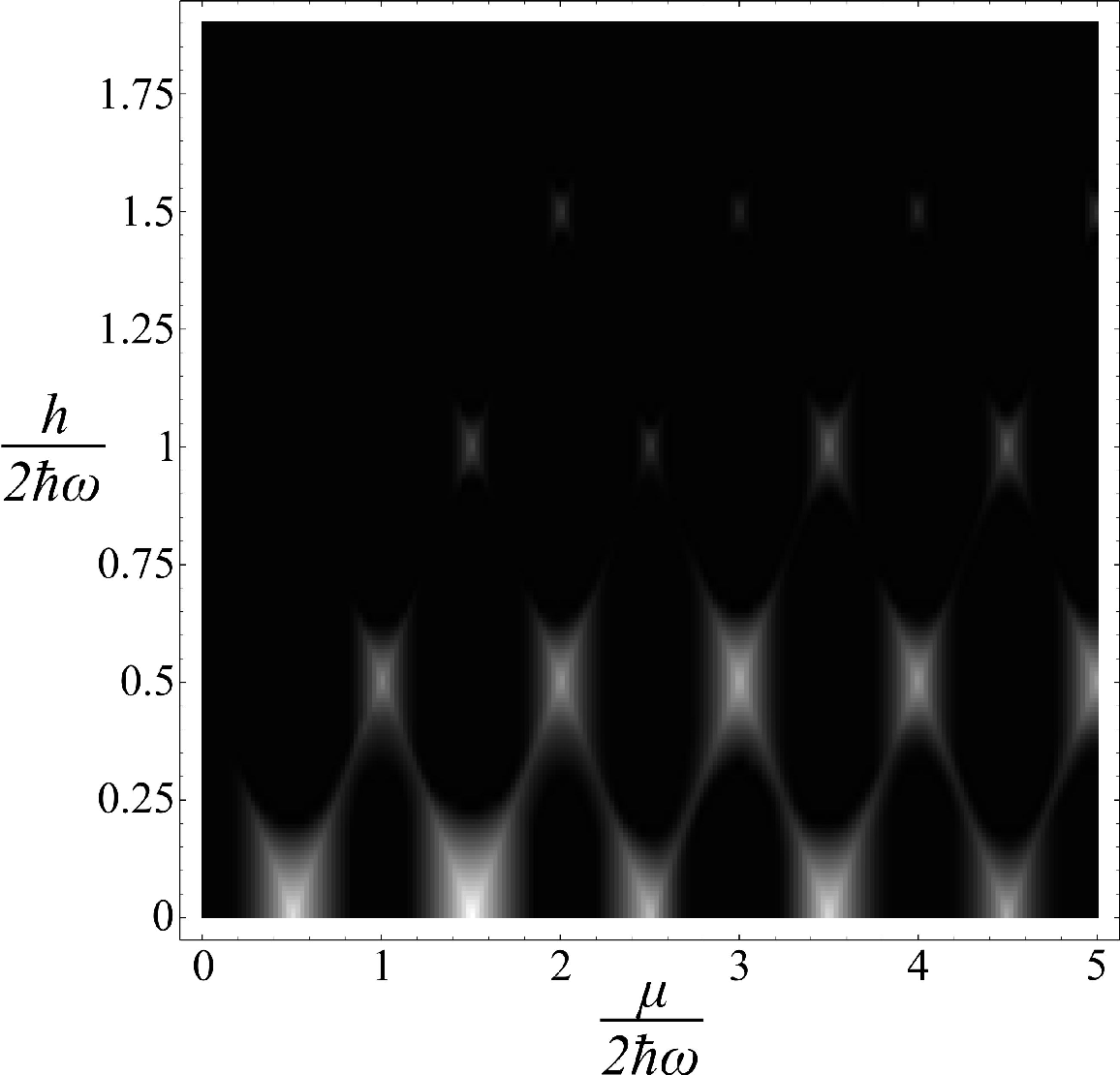}}
\caption{\label{PairingEnergy}Density plot of the pairing free energy density $\mathcal{F}(\phi_{n,l})-\mathcal{F}(0)$ at (a) unitarity and (b) $\nu=0.15$. Brightness indicates the amount of pairing.}
\end{figure*}

Re-entrant superfluidity is reflected in the robust paired states which occur whenever the chemical potentials of the two atom species, $\mu \pm h$, approach Landau levels. This is best demonstrated in Figure \ref{PairingEnergy} which shows pairing free energy. As a result, a sequence of paired-state islands extends to arbitrary large $h$ and survives at $h=0$ regardless of how weak the attractive interactions between atoms are. Fig.\ref{PhaseDiagram} and Fig.\ref{PhaseDiagramFFLO} reveal a fundamental relationship between $n \ge 1$ Landau level FFLO states and re-entrant pairing in the deep quantum limit.

\begin{figure}
\centering
\includegraphics[width=1.1in]{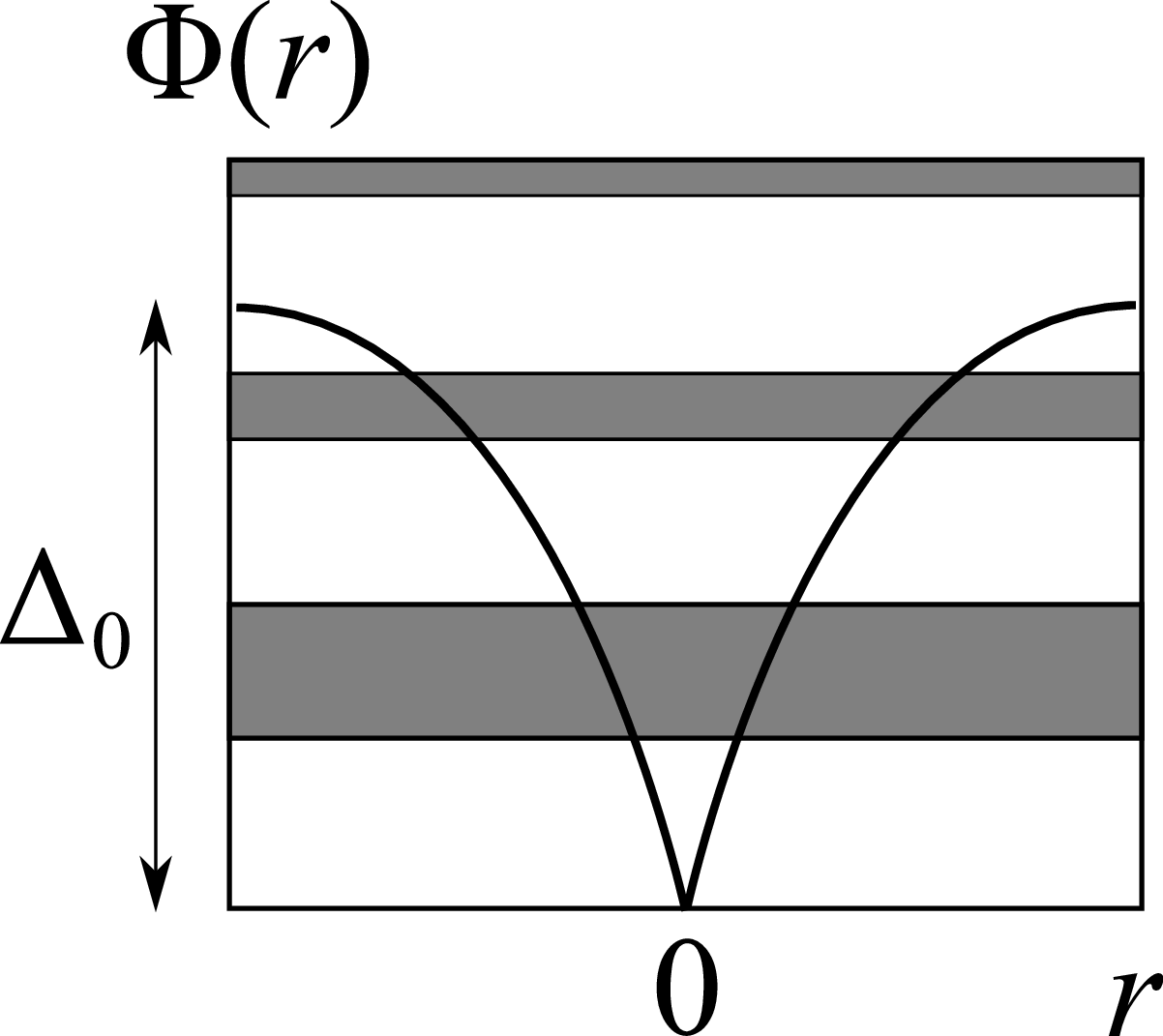}
\caption{\label{VortexCore}Vortex core bands in relation to the superfluid gap near a vortex. The quasiparticle spectrum reduces to nearly-degenerate ``Landau levels'' at high energies, but resembles a band-structure shaped by the vortex lattice periodic potential at low energies, especially below the pairing scale $|\varphi|$.}
\end{figure}

Superfluid states polarized by Zeeman field, such as FFLO states, are elusive and difficult to stabilize except in the presence of a vortex lattice. Isolated vortices in $s$-wave superfluids always have quasiparticle states localized in the core \cite{Sensarma2006}, so that the core bands in a vortex lattice, broadened by quantum tunneling between cores, naturally lie below the energy associated with the pairing gap scale (see Fig.\ref{VortexCore}). The Zeeman field needed to inject quasiparticles into these bands is lower than the pairing gap and hence well below the Pauli-Clogston limit for pair-breaking. Therefore, Zeeman field is initially unable to destroy the superfluid despite injecting spin. The stability of polarized vortex lattices goes beyond this picture and even Zeeman fields larger than the pairing gap are often not enough to destroy the superfluid (for example, there are second-order transitions between FFLO states and integer quantum Hall insulators). This is not difficult to understand having in mind that the superfluid is full of ``holes'' arranged in a lattice which can absorb injected spin. The matter is completely trivialized in three dimensions where vortex cores are always polarized by any finite Zeeman field. In two-dimensions, however, it takes a finite $h$ and a ``metal-insulator'' transition, which at least in some cases is first order, to polarize the cores.

The spin-polarized vortex lattices are not considered special unless the vortex lattice structure and the number of singularities change due to spin injection. Vortex lattice FFLO states contain additional vortex-antivortex pairs in every unit-cell, which is favored at large $h$ by the virtue of providing more regions in space where superfluidity is suppressed and injected spin absorption is easier. FFLO order parameters are condensates with dominant amplitude in a bosonic Landau level $n\ge 1$. This kind of FFLO states is also relatively stable for the same reason as the plain polarized vortex lattices in $n=0$ condensates, although undoubtedly more difficult to realize.

It should be pointed out that the distinction between FFLO states and the plain spin-polarized vortex lattices is somewhat artificial and aims to merely separate unusual from common phenomena. In terms of symmetries, which are the main criteria for identifying phases in condensed matter physics, there is no distinction between FFLO and plain polarized vortex lattices. Both kinds have spontaneously broken translational and rotational symmetries by the superfluid order parameter and spin polarization. Also, both kinds come in two varieties, with a Fermi sea or band-insulator of polarized quasiparticles. The formal difference between the two kinds comes from the dominant bosonic Landau level $n$ in the condensate, but this in its own right does not affect symmetry properties as demonstrated in the Fig.\ref{VortexLattice}. The condensate Landau level content defines the internal structure of a vortex lattice unit-cell rather than its size or shape. This can be appreciated even further by realizing that deep in the superfluid state the condensate must acquire amplitudes at multiple Landau levels, without changes of symmetries, in order to accommodate the evolution of the vortex core structure (for example, the growth of cores in the BCS limit). Condensates in a single Landau level strictly exist only at the second order superfluid-insulator transitions.

\begin{figure}
\centering
\includegraphics[width=2.3in]{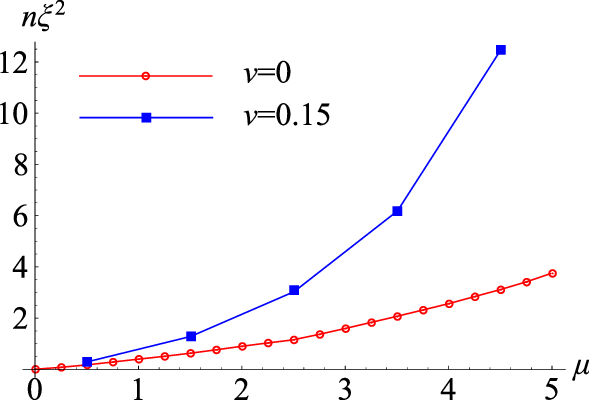}
\caption{\label{XiPrm}(color online) Parameter $n\xi^2$ at $h=0$. Saddle point approximation is quantitatively justified if $n\xi^2\gg 1$.}
\end{figure}

The limitations of saddle point approximation (SPA), which was used to calculate the phase diagrams, are discussed in \cite{nikolic:144507}. An indicator for the quantitative applicability of SPA is $n\xi^2\gg 1$, where $n$ is atom density and $\xi\sim\hbar\sqrt{2\mu/m|\varphi|^2}$ the BCS coherence length. $\xi$ grows rapidly with density or detuning as shown in Fig.\ref{XiPrm}. The stability of integer quantum Hall insulators against fluctuations is guaranteed in the mean-field regimes when $\xi$ is large, while otherwise pronounced instabilities in the particle-hole channel can occur (for example, as found in Ref.\cite{Tesanovic1994, Moller2007}).

The only qualitative shortcoming of SPA reflected in the phase diagram are the missed ``vortex liquid'' phases which can be viewed as strongly correlated insulators of Cooper pairs with non-universal properties. Starting in a superfluid state, quantum fluctuations can lead to first order vortex lattice melting and result with a number of different phases, including density waves and fractional quantum Hall states of Cooper pairs (with even-denominator filling factor). These phases inevitably mask all second order superfluid-insulator transitions shown in Fig.\ref{PhaseDiagram} and grow at the expense of the vortex lattice regions, especially at the shown low densities when the number of atoms per vortex is of the order of one. The SPA still qualitatively captures the boundary between the paired correlated insulators and unpaired atomic quantum Hall states, and indirectly predicts the existence of metallic and insulating spin-polarized vortex liquids formed in various bosonic Landau levels. The $T=0$ ``metal-insulator'' transitions at finite $h$ are expected to survive within vortex liquid states as a remnant of broadened Landau levels. We conclude with a note that vortex liquids may be responsible for some unconventional properties of high-temperature superconductors \cite{Ong2007}.

\begin{figure}
\subfigure[{}]{\includegraphics[height=1.1in]{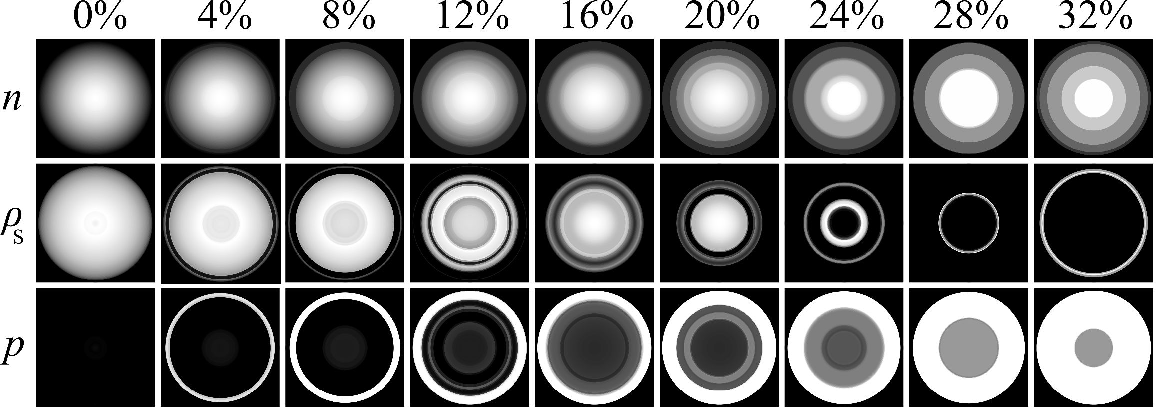}}
\subfigure[{}]{\includegraphics[height=1.1in]{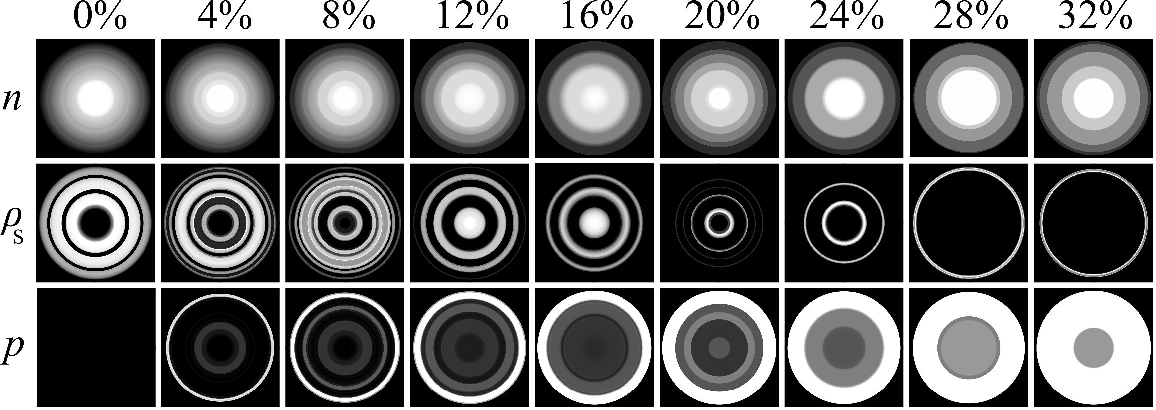}}
\caption{\label{Trap}Radial $(r)$ profiles of total particle density $n(r)$, Cooper pair density $\rho_s(r)$ and polarization $p(r)$ as a function of total trap polarization $P=(N_\uparrow-N_\downarrow)/N$. The total particle number $N=N_\uparrow+N_\downarrow$ in the trap is $N=3.5\cdot 10^5$; dimensionless detuning is (a) $\nu=0$, (b) $\nu=0.15$. The total polarization $P$ is indicated on the top. These plots illustrate sharp density features at very low temperatures. Starting from $P=0$ the initial trend is that the superfluid core shrinks as the added ``majority-spin'' atoms are pushed to the trap boundary. However, the ring structure quickly acquires complexity from the Fig.\ref{PhaseDiagram}: moving from $r=0$ toward the trap edge follows a trajectory of reducing $\mu$ while keeping $h$ constant in Fig.\ref{PhaseDiagram}. Characteristic superfluid rings separated by normal regions are a signature of re-entrant superfluidity, and polarization can be finite in a superfluid ring (FFLO states, or SF-I and SF-M in Fig.\ref{PhaseDiagram}). The distinction between SF-I and SF-M phases can be made by whether $p(r)$ is a quantized constant or smoothly varying respectively. Superfluid regions replaced by vortex liquids are expected to have strong pairing fluctuations \cite{Partridge2005, Schunck07}. A currently experimentally unreachable trap harmonic frequency $\omega_\perp = (1+5\cdot 10^{-5})\omega$ was chosen to justify local density approximation, and the shown spatial region is $r<500 \sqrt{\hbar/(2m\omega)}$. All other parameters are comparable with already accomplished experimental setups \cite{Stock2005, Schweikhard2004}.}
\end{figure}

Temperatures needed to explore the vortex-FFLO phase diagram in cold atom experiments should be below $T^* = k_B^{-1} \hbar\omega$, which is a few nK for typical $\omega\sim 2\pi\times 100$ Hz and within reach by evaporative cooling. Critical temperatures for superfluidity are generally larger, being of the order of the $T=0$ pairing gap shown in Fig.\ref{phdiag3d}(a) and Fig.\ref{PairingGap}. The main difficulty would be working with a low enough density of trapped atoms, ideally not much larger than $n_0 = m\omega / (\pi\hbar) \sim 10^6 - 10^7$ cm$^{-2}$, the density for a single populated Landau level \cite{Stock2005, Schweikhard2004}. New experimental techniques may be able to overcome this difficulty in the future \cite{Gunter2009, Spielman2009}. If the deep quantum limit becomes achievable in fast-rotating cold atom gases, time-of-flight images could reveal intricate ring-shaped density, polarization and pairing profiles shown in Fig.\ref{Trap}. In general, alternating rings of different superfluid and insulating states can be expected as the effective radius-dependent chemical potential crosses a sequence of Landau levels, a phenomenon known as ``quantum oscillations'' in condensed matter physics.

The spatial profiles in Fig.\ref{Trap} are determined by a naive application of local density approximation (LDA). This can be valid only in the parts of the trap where the spatial variations of local system properties, such as the superfluid order parameter, are very gradual. For example, a superfluid ring should contain a significant number of vortices across its thickness in order to be resolved, so the intricate features near the trap boundaries are especially jeopardized. Unfortunately, the rotation rate required to justify LDA in the deep quantum limit would need to be extremely close to the trapping harmonic frequency and currently is beyond experimental reach. However, the main features of the phase diagram in Fig.\ref{PhaseDiagram} are expected to persist at much higher densities than shown, so that achieving the deep quantum limit may not be necessary. Spatial resolving of the characteristic alternating rings is expected to be even harder if many fermionic Landau levels are populated. Instead, quantum oscillations could be sought in the oscillatory dependence of the central disk radius (and nearby ring radii) on rotation rate, or the harmonic trap frequency. Due to the parabolic trap potential, spatial features near the trap center are stretched with respect to those near the trap boundaries, and fine oscillations might be observable.

\section{Diagonalization of the Bogoliubov-de Gennes Hamiltonian}

A typical mean-field approximation applied to fermionic systems starts from a linearized Bogoliubov-de Gennes Hamiltonian in the light of the fact that fermions with energies far removed from the Fermi level do not contribute much to dynamics. Linearization greatly simplifies solving gap equations, especially when the order parameter has multiple components like in our case. However, linearization is problematic in the present case because the order parameter also describes singularities associated with vortices. This issue was extensively discussed in Ref.~\cite{Melikyan2006} and handled by a sophisticated mathematical machinery of self-adjoint representations.

Here, we take a different approach, conceptually simpler and straight-forward to implement numerically. Instead of linearizing the BdG Hamiltonian, we diagonalize it exactly for any given order parameter, and then substitute the obtained spectrum in the expression for mean-field free energy density. Minimizing free energy with respect to the order parameter is in the worst case equivalent to solving gap equations, but generally better because unstable and meta-stable solutions are automatically discarded.

We avoid using real-space representation and instead represent vortices and all particle states in the basis of Landau levels, using Landau gauge $\Av = - 2m\omega y \xh$. The Landau levels in the absence of interactions are $\varepsilon_{n} = 2\hbar\omega \left( n + \frac{1}{2} \right) - \mu$, and all fermion and boson wavefunctions are expanded as:
\begin{eqnarray}\label{WF2}
\psi_{\alpha}(\rv) & = & \sum_n \int \frac{\dd k_x}{2\pi} \varphi_n \left(
  \rv \sqrt{\frac{2m\omega}{\hbar}} ~ ; ~ \frac{k_x}{\sqrt{2m\hbar\omega}} \right) \psi_{\alpha,n,k_x} \nonumber \\
\Phi(\rv) & = & \sum_n \int \frac{\dd q_x}{2\pi} 2^{\frac{1}{4}} \varphi_n \left(
  \rv \sqrt{\frac{4m\omega}{\hbar}} ~ ; ~ \frac{q_x}{\sqrt{4m\hbar\omega}} \right) \phi_{n,q_x} \ , \nonumber
\end{eqnarray}
with $\varphi_n(\Bf{\xi};\eta)$ given by ~(\ref{WF}). The quantum numbers in the Landau gauge are momentum $k_x$ and Landau-level index $n$.

The short-range pairing interaction between fermions takes the following form:
\begin{eqnarray}
& & \int\dd^2r \left\lbrack \Phi(\rv) \psi^{\dagger}_{\uparrow}(\rv) \psi^{\dagger}_{\downarrow}(\rv) + h.c. \right\rbrack =
\\ & & ~~~  = \sum_n \sum_{m_1,m_2} \int \frac{k_x}{2\pi} \frac{q_x}{2\pi} \Biggl\lbrack
    \Gamma_{m_1,m_2}^n \left( \frac{k_x}{\sqrt{2m\hbar\omega}} \right) \times \nonumber
\\ & & ~~~~ \times \phi^{\phantom{\dagger}}_{n,q_x}
    \psi^{\dagger}_{\uparrow,m_1,k_x+\frac{q_x}{2}} \psi^{\dagger}_{\downarrow,m_2,-k_x+\frac{q_x}{2}} + h.c. \Biggr\rbrack
    \ , \nonumber
\end{eqnarray}
where $\Gamma$ is the bare vertex function:
\begin{eqnarray}\label{Vertex1}
&& \Gamma_{m_1,m_2}^n(\eta) = \frac{2^{-(n+m_1+m_2)/2}}{\sqrt{\pi n!m_1!m_2!}}
  \left(\frac{2}{\pi}\right)^{\frac{1}{4}} e^{-\eta^2} \times \\
&&  \times \int\limits_{-\infty}^{\infty} \dd\xi_y e^{-2\xi_y^2} H_n(\sqrt{2}\xi_y) H_{m_1}(\xi_y+\eta) H_{m_2}(\xi_y-\eta)
  \nonumber \ .
\end{eqnarray}
Consequently, the BdG Hamiltonian looks highly non-diagonal in Landau level representation for a generic order parameter. The block which couples the Nambu states $\la \psi_{\uparrow, m_1, k_{x1}}^{\dagger} , \psi_{\downarrow, m_1, k_{x1}}^{\phantom{\dagger}}\vert$ and $\vert \psi_{\uparrow, m_2, k_{x2}}^{\phantom{\dagger}} , \psi_{\downarrow, m_2, k_{x2}}^{\dagger}\rangle$ is:
\begin{eqnarray}\label{BdG1}
&& H_{m_1,m_2}^{\Tr{BdG}}(k_{x1},k_{x2}) = \\
&& ~~~~ \left( \begin{array}{cc}
  \varepsilon_{m_1}\delta_{m_1,k_{x1};m_2,k_{x2}} &
    \Delta_{m_1,m_2}(k_{x1},k_{x2}) \\
  \overline{\Delta}_{m_1,m_2}(k_{x1},k_{x2}) &
    -\varepsilon_{m_2}\delta_{m_1,k_{x1};m_2,k_{x2}}
\end{array} \right) \nonumber \ ,
\end{eqnarray}
where $\delta_{m_1,k_{x1};m_2,k_{x2}} = \delta_{m_1,m_2}2\pi\delta(k_{x1}-k_{x2})$ and
\begin{eqnarray}\label{BdGDelta1}
&& \Delta^{\phantom{*}}_{m_1,m_2}(k_{x1},k_{x2}) = \overline{\Delta}^*_{m_2,m_1}(k_{x2},k_{x1}) = \\
&& ~~~~~~ = \sum_n \phi^{\phantom{*}}_{n,k_{x1}+k_{x2}} \Gamma_{m_1,m_2}^{n}\left(\frac{k_{x1}-k_{x2}}{2\sqrt{2m\hbar\omega}}\right) \ . \nonumber
\end{eqnarray}

However, the order parameter must have a certain degree of periodicity in order to capture a vortex lattice. A finite density of vortices cannot be described by a truly periodic function due to the presence of circulating supercurrents. On the other hand, superfluid density must be periodic. In order to satisfy this requirement and simultaneously obtain the phase of the order parameter which properly winds by $2\pi$ around each vortex we construct a superposition of Landau-level states with quasi-periodic properties. The order parameter function (\ref{OP}) is periodic in $x$-direction by the virtue of being a superposition of plane-waves at discrete momenta $q_x = (l+n_vj)\delta q$. Since every Landau level eigenfunction in Landau gauge is displaced in $y$-direction by an amount proportional to $q_x$, we obtain ``periodicity'' in $y$-direction as well ($n_v$ allows freedom to independently set periodicity in $x$ and $y$ directions). However, $\Phi(\rv)$ is periodic in $y$-direction only up to a phase shift, which is naturally inherited from the Landau level wavefunctions and necessary to describe vorticity.

The amplitudes $\phi_{n,l}$ in ~(\ref{OP}) for $n\ge 0$, $n_v>l\ge 0$ are variable complex numbers which determine the structure of the vortex lattice in some complicated manner. Since only a discrete set of momenta is populated by the order parameter, we can use Bloch's theorem to simplify the BdG Hamiltonian. Below we relabel the fermion momentum $k_x \to k_x + \lambda \delta q$, where the new label $k_x\in(-\delta q/2,\delta q/2)$ is a conserved ``crystal momentum'' and $\lambda$ is an integer. The BdG Hamiltonian is diagonal in $k_x$ and the block which couples the Nambu states $\la \psi_{\uparrow, m_1, k_x+\lambda_1\delta q}^{\dagger} , \psi_{\downarrow ,m_1, -k_x+\lambda_1\delta q}^{\phantom{\dagger}}\vert$ and $\vert \psi_{\uparrow,m_2, k_x+\lambda_2\delta q}^{\phantom{\dagger}} , \psi_{\downarrow, m_2, -k_x+\lambda_2\delta q}^{\dagger}\rangle$ becomes:
\begin{eqnarray}\label{BdG2}
&& H_{m_1,\lambda_1;m_2,\lambda_2}^{\Tr{BdG}}(k_x) = \\
&& ~~~~ = \left(
\begin{array}{cc}
  \varepsilon_{m_1}\delta_{m_1,m_2}\delta_{\lambda_1,\lambda_2} &
    \Delta_{m_1,\lambda_1;m_2,\lambda_2}(k_x) \\
  \overline{\Delta}_{m_1,\lambda_1;m_2,\lambda_2}(k_x) &
    -\varepsilon_{m_2}\delta_{m_1,m_2}\delta_{\lambda_1,\lambda_2}
\end{array}
\right) \nonumber
\end{eqnarray}
where
\begin{eqnarray}\label{BdGDelta2}
&& \Delta^{\phantom{*}}_{m_1,\lambda_1;m_2,\lambda_2}(k_x) = \overline{\Delta}^*_{m_2,\lambda_2;m_1,\lambda_1}(k_x) = \\
&& ~~~~~~ = \sum_{n,l} \phi^{\phantom{*}}_{n,l} \Gamma_{m_1,m_2}^{n}\left(\frac{2k_x+(\lambda_1-\lambda_2)\delta q}
    {2\sqrt{2m\hbar\omega}}\right) \ . \nonumber
\end{eqnarray}
Momentum conservation $l-(\lambda_1+\lambda_2) = 0 (\Tr{mod}~n_v)$ is implicitly assumed in the sum over $l$. This unusual constraint comes from the fact that two fermions carry momenta $k_x+\lambda_1\delta q$ and $-k_x+\lambda_2\delta q$, while a boson can carry momentum $(l + n_vj)\delta q$, and $j$ is summed over all integers. Due to this constraint, the gap functions $\Delta$ and $\overline{\Delta}$ are not simply functions of $\lambda_1-\lambda_2$. If we define $\lambda_i=\delta\lambda_i+n_vL_i$, where $L_i$ is any integer, and $n_v>\delta\lambda_i\ge 0$, then the gap amplitudes depend on $\delta\lambda_1$, $\delta\lambda_2$ and $L_1-L_2$. Now we can introduce a variable $\theta\in(-\pi,\pi)$ canonically conjugate to $L$ and diagonalize the Hamiltonian up to a finite non-diagonal representation:
\begin{eqnarray}\label{BdG3}
&& H_{m_1,\delta\lambda_1;m_2,\delta\lambda_2}^{\Tr{BdG}}(k_x,\theta) = \\
&& \left(
\begin{array}{cc}
  \varepsilon_{m_1}\delta_{m_1,m_2}\delta_{\delta\lambda_1,\delta\lambda_2} &
    \Delta_{m_1,\delta\lambda_1;m_2,\delta\lambda_2}(k_x,\theta) \\
  \overline{\Delta}_{m_1,\delta\lambda_1;m_2,\delta\lambda_2}(k_x,\theta) &
    -\varepsilon_{m_2}\delta_{m_1,m_2}\delta_{\delta\lambda_1,\delta\lambda_2}
\end{array}
\right) \nonumber
\end{eqnarray}
where
\begin{eqnarray}\label{BdGDelta3}
&& \Delta^{\phantom{*}}_{m_1,\delta\lambda_1;m_2,\delta\lambda_2}(k_x,\theta) =
   \overline{\Delta}^*_{m_2,\delta\lambda_2;m_1,\delta\lambda_1}(k_x,\theta) = \\
&& = \sum_{n,l} \phi^{\phantom{*}}_{n,l} \delta_{(\delta\lambda_1+\delta\lambda_2-l)(\Tr{mod}~n_v),0} \nonumber \\
&& \times \sum_{\Delta L} e^{-i\Delta L \theta}
   \Gamma_{m_1,m_2}^{n}\left(\frac{2k_x+(n_v\Delta L + \delta\lambda_1-\delta\lambda_2)\delta q}
    {2\sqrt{2m\hbar\omega}}\right) \nonumber
\end{eqnarray}
The conserved quantum numbers $k_x$ and $\theta$ are physically similar to ``crystal momenta'' in $x$ and $y$ directions shaped by the vortex lattice. Broadened Landau levels are dispersive with respect to these quantum numbers, a manifestation of degeneracy lifting by pairing.

The Hamiltonian ~(\ref{BdG3}) can be easily diagonalized numerically because all continuous parameters are good quantum numbers which live in a finite range, and all remaining parameters are discrete. The numerical calculation of ~(\ref{BdGDelta3}) can be done very accurately since the contributions of large $\Delta L$ are exponentially suppressed by the vertex functions.

\section{Quantum fluctuations}\label{QFluct}

Here we describe a systematic perturbative framework for calculating fluctuation corrections and then discuss fluctuation effects at the qualitative level. Pursuing these corrections numerically is beyond the scope of this paper, as we aim to reveal only the fundamental features of the phase diagram.

A complete quantum field theory which describes interacting fermions in the unitarity regime is given by the following imaginary-time action (in units $\hbar=1$):
\begin{eqnarray}
&& S = \int\dd\tau\dd^dr \Bigl\lbrack \psi_{i\alpha}^\dagger \left( \frac{\partial}{\partial\tau}
     - \frac{(-i\nv-\Av)^2}{2m} - \mu \right) \psi_{i\alpha}^{\phantom{\dagger}} \nonumber \\
&& + h \left( \psi_{i\uparrow}^\dagger \psi_{i\uparrow}^{\phantom{\dagger}} -
               \psi_{i\downarrow}^\dagger \psi_{i\downarrow}^{\phantom{\dagger}} \right)
    + \Phi^\dagger \psi_{i\uparrow}^{\phantom{\dagger}} \psi_{i\downarrow}^{\phantom{\dagger}}
    + \Phi \psi_{i\downarrow}^\dagger \psi_{i\uparrow}^\dagger \Bigr\rbrack
    + S_{\Tr{reg}} \nonumber
\end{eqnarray}
The fermionic Grassmann fields $\psi_{i\alpha}$ carry spin $\alpha\in\lbrace\uparrow,\downarrow\rbrace$ and come in $i=1\dots N$ flavors. This endows the model with SP($2N$) symmetry; the physical case of interest is $N=1$. The part $S_{\Tr{reg}}$ contains regularization and an effective chemical potential for the $\Phi$ field (detuning from the Feshbach resonance) \cite{nikolic:144507}.

The reason for generalizing the physical problem to large $N$ is only to provide a mathematical limit in which the approximations become exact despite the large interaction coupling. Fermions can be integrated out exactly to obtain an effective bosonic action which can be written graphically using Feynman diagrams:
\begin{eqnarray}\label{Action2}
&& S_{\Tr{eff}} = S'_{\Tr{reg}} + \\[-0.4cm]
&& ~ \raisebox{-0.3cm}{\includegraphics[height=0.8cm]{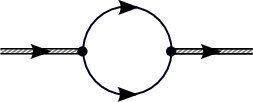}} +
  \frac{1}{2} \raisebox{-0.63cm}{\includegraphics[height=1.5cm]{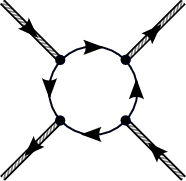}} +
  \frac{1}{3} \raisebox{-0.82cm}{\includegraphics[height=1.8cm]{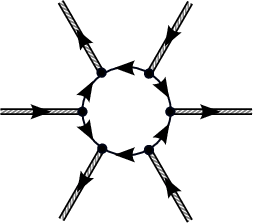}}
  + \cdots \ . \nonumber
\end{eqnarray}
Double lines represent $\Phi$ and $\Phi^\dagger$, depending on the direction of arrow with respect to vertex, while the solid lines are fermions. Now, the field $\Phi$ obtains dispersion in the first bubble diagram and represents bosonic degrees of freedom, Cooper pairs or molecules. The remaining terms are vertices for interactions between bosons. Each fermion loop carries a factor of $N$, so that in the large-$N$ limit bosonic fluctuations are suppressed. The effective boson propagator is proportional to $1/N$ and perturbation theory produces $1/N$ expansions for all thermodynamic functions.

The conceptually simplest way to include quantum fluctuations in the presence of a condensate is to take the $1/N$ expansion in the representation which diagonalizes the BdG Hamiltonian. Then, the fermion lines represent the BdG quasiparticle Green's function
\begin{equation}
G_n(i\omega) = \frac{1}{-i\omega+E_{\Bf n}(\Phi)}
\end{equation}
where $E_{\Bf n}(\Phi)$ are BdG eigenvalues, and ${\Bf n}$ are appropriate quantum numbers. There is no use of ``anomalous'' propagators in this approach, but the vertex function obtains a more complicated form.

The $1/N$ expansion is semi-classical in character and allows systematic calculation of fluctuation corrections to any observable. This unique ability to maintain controlled approximations in a strongly-interacting theory is what motivates the use of an Sp($2N$) theory. In the limit of large $N$ the mean-field phase diagram discussed in the first section is guarantied to be qualitatively correct, with two notable caveats which will be mentioned shortly. However, certain effects in physical systems ($N\sim 1$) are difficult to capture through fluctuation corrections in this theory. Most notably, they include instabilities in particle-hole channel, the appearance of insulators with density-wave order. While different types of mean-field theory (Hartree-Fock for example) are better suited for finding such ordered states, they do not allow a systematic treatment of fluctuations, making it conceptually difficult to explore the competition between different orders and their stability against fluctuations.

There are a couple of known fluctuation effects which do not follow in a straight-forward manner from $1/N$ expansions, and qualitatively affect the phase diagram regardless of how large $N$ is. First, it has been argued that the two-dimensional superfluid hosting a vortex lattice is not a condensate even at zero temperature (no single-particle bosonic state is macroscopically populated) \cite{Moore1989, Tesanovic1994, Sinova2002}. Namely, quantum zero-point motion of vortices, or would be Goldstone modes in a hypothetical condensate, restore U(1) symmetry down to algebraic correlations. The correlation length remains formally infinite, but the only surviving long-range order is the broken translational symmetry by the vortex lattice. This implies that the order-parameter ~(\ref{OP}) is ultimately not an adequate characterization of the ground-state. However, as discussed in Ref.\cite{nikolic:144507}, the mean-field theory based on a condensate order parameter is useful at sufficiently short length and time scales. Many experimental observables, such as particle and spin density, the presence of vortices and their lattice structure, are local properties and can be qualitatively captured even by the mean-field theory, especially since the trapped cold atoms have very slow dynamics on the measurement scales.

Another important phenomenon, discussed in the section \ref{secMain}, is the existence of ``vortex liquids''. These strongly correlated insulators of Cooper pairs are entirely missed by the saddle-point phase diagram in Fig.~\ref{PhaseDiagram}, but the amount of space they take scales as $1/N$ in the large-$N$ limit. Since this limit suppresses quantum fluctuations of bosonic degrees of freedom, it is natural to expect that the effects of vortex fluctuations are also weak deep in the BCS limit. Near unitarity and in the BEC limit the physical systems will be affected by vortex quantum motion.

\section{Conclusions}

In this paper we analyzed the phase diagram of spin-polarized fermionic superfluids and quantum Hall insulators in two-dimensions at $T=0$, and discussed observable consequences at low temperatures in cold atom experiments. We hope that the richness of the phase diagram will stimulate further research, especially experiments near unitarity which could provide valuable answers to long standing questions about the stability of superconducting phases in large magnetic fields and the nature of strongly correlated normal states.

We emphasized that FFLO states in vortex lattices are much more stable than their counterparts in static uniform systems. This is a simple consequence of the fact that vortex cores can absorb injected spin relatively easily because of locally suppressed superfluidity. Even though reaching deep quantum limit is a great experimental challenge, the unusual physics of rotated imbalanced Fermi gases may be observed through ``quantum oscillation'' measurements.

I am very grateful to Randal Hulet, Zoran Hadzibabic and Leslie Baksmaty for useful comments and discussions. Numerical calculations were performed on Rice University computing clusters. This research was supported by W.~M.~Keck Program in Quantum Materials.

%\bibliographystyle{apsrev}
%\bibliography{/home/dasko/Science/Bibliography/references}

\end{document}